\documentclass[transmag]{IEEEtran}
\IEEEoverridecommandlockouts 
\usepackage{amsmath,graphicx}
\usepackage{cite}
\usepackage[numbers,sort&compress]{natbib}
\usepackage{float}
\usepackage{tabularx} % in the preamble
\usepackage{balance}
\usepackage{multirow}
\usepackage{makecell}
\usepackage{pgfplots}
\usepackage{amsmath,amssymb,amsfonts}
\usepackage{algorithmic}
\usepackage{graphicx}
\usepackage{xcolor}
\usepackage{boldline, soul}
\usepackage{booktabs}
\usepackage{indentfirst}
\usepackage{caption}
\usepackage{amssymb}
\usepackage{amsfonts}
\usepgfplotslibrary{patchplots}

\newcolumntype{P}[1]{>{\centering\arraybackslash}p{#1}}
\newcolumntype{M}[1]{>{\centering\arraybackslash}m{#1}}
\newcolumntype{C}[1]{>{\raggedright\arraybackslash}m{#1}}

\usepackage{pifont}% http://ctan.org/pkg/pifont
\newcommand{\cmark}{\ding{51}}%
\newcommand{\xmark}{\ding{55}}%

\usepackage{amsmath,amssymb,amsfonts}
\usepackage{algorithmic}
\usepackage{graphicx}
\usepackage{textcomp}
\usepackage{xcolor}
\def\BibTeX{{\rm B\kern-.05em{\sc i\kern-.025em b}\kern-.08em
    T\kern-.1667em\lower.7ex\hbox{E}\kern-.125emX}}
    
    \setcellgapes{3pt}
\def\BibTeX{{\rm B\kern-.05em{\sc i\kern-.025em b}\kern-.08em
    T\kern-.1667em\lower.7ex\hbox{E}\kern-.125emX}}

\pgfplotsset{compat=1.13}

\makeatletter
\newcommand{\thickhline}{%
    \noalign {\ifnum 0=`}\fi \hrule height 1pt
    \futurelet \reserved@a \@xhline
}
\newcolumntype{"}{@{\hskip\tabcolsep\vrule width 1pt\hskip\tabcolsep}}

\usepackage{makecell}
\usepackage{multirow}
\usepackage{booktabs}
\usepackage{pdfpages}
\usepackage[acronym]{glossaries}

\makeglossaries
\newacronym{cnn}{CNN}{Convolutional Neural Network}
\newacronym{vvc}{VVC}{Versatile Video Coding}
\newacronym{hevc}{HEVC}{High Efficiency Video Coding}
\newacronym{dbf}{DBF}{De-Blocking filter}
\newacronym{qp}{QP}{Quantization Parameter}
\newacronym{sao}{SAO}{Sample Adaptive Offset}
\newacronym{alf}{ALF}{Adaptive Loop Filter}
\newacronym{sr}{SR}{Super Resolution}
\newacronym{pp}{PP}{Post Processing}
\newacronym{ilf}{ILF}{In-Loop Filtering}
\newacronym{qe}{QE}{Quality Enhancement}
\newacronym{gop}{GoP}{Group of Pictures}
\newacronym{evc}{EVC}{Essential Video Coding}
\newacronym{ctu}{CTU}{Coding Tree Unit}
\newacronym{mse}{MSE}{Mean Squared Error}
\newacronym{cu}{CU}{Coding Unit}
\newacronym{ipm}{IPM} {Intra Prediction Modes}
\newacronym{tu}{TU}{Transform Unit}
\newacronym{pu}{PU}{Prediction Unit}
\newacronym{msssim}{MS-SSIM}{Multi-Scale Structural SIMilarity}
\newacronym{pqf}{PQF}{Peak Quality Frame}
\newacronym{mc}{MC}{Motion Compensated}
\newacronym{gpu}{GPU}{Graphic Processing Unit}
\newacronym{ai}{AI}{All Intra}
\newacronym{ra}{RA}{Random Access}
\newacronym{ld}{LD}{Low Delay}
\newacronym{vr}{VR}{Virtual Reality}
\newacronym{dc}{DC}{direct current}
\newacronym{bd-br}{BD-BR}{Bj\o ntegaard Delta Bit Rate}
\newacronym{ct}{CT}{Coding Type} 
\newacronym{rt}{RT}{Run-Time}

\newacronym{mv}{MV}{Motion Vector}
\newacronym{psnr}{PSNR}{Peak Signal-to-Noise Ratio}

\newacronym{ctc}{CTC}{Common Test Condition} 
\newacronym{tum}{TUM}{Transform Units Map}
\newacronym{pm}{PM}{Partitioning Map}
\newacronym{qm}{QM}{QP Map}
\newacronym{cum}{CUM}{Coding Units Map}
\newacronym{resm}{ResM}{Residual Map}
\newacronym{predm}{PrM}{Prediction Map}
\newacronym{mm}{MM}{Mean Map}
\newacronym{ipmm}{IPMM}{Intra Prediction Mode Map}

\newacronym{dcad}{DCAD}{Deep CNN-based Auto Decoder}
\newacronym{IFCNN}{IFCNN}{In-Loop filter CNN }
\newacronym{STResNet}{STResNet}{Spatial-Temporal Residue Network}
\newacronym{DCAD}{DCAD}{Deep CNN-based Auto Encoder}
\newacronym{DSCNN}{DSCNN}{Decoder-side Scalable Convolutional Neural Network}
\newacronym{MMS-net}{MMS-net}{Multi-modal/Multi-scale Convolutional Neural Network}
\newacronym{VRCNN}{VRCNN}{Variable-filter-size Residue-learning CNN}
\newacronym{MSDD}{MSDD}{Multi-Scale deep decoder}
\newacronym{QECNN}{QECNN}{Quality Enhancement Convolutional Neural Network}
\newacronym{MFQE}{MFQE}{Multi-Frame Quality Enhancement }
\newacronym{CNNF}{CNNF}{CNN Filter}
\newacronym{RHCNN}{RHCNN}{Residual Highway CNN}
\newacronym{FECNN}{FECNN}{Frame Enhancement CNN}
\newacronym{Residual-VRN}{Residual-VRN}{Residual-based Video Restoration Network}
\newacronym{MGANet}{MGANet}{Multi-frame Guided Attention Network}
\newacronym{MLSDRN}{MLSDRN}{Multi-Channel Long-Short term Dependency Residual Network}
\newacronym{ADCNN}{ADCNN}{Attention based Dual-scale CNN}
\newacronym{MM-CU-PRN}{MM-CU-PRN}{Multi-scale Mean value of CU-Progressive Rethinking Network}
\newacronym{SDTS}{SDTS}{Spatial Details and Temporal Structure}
\newacronym{VRCNN-BN}{VRCNN-BN}{Variable Filter-size Residue-learning CNN with batch normalization layer}
\newacronym{MIF}{MIF}{Multi-frame In-loop Filter}
\newacronym{MRRN}{MRRN}{Multi-Reconstruction Recurrent Residual Network}
\newacronym{RRCNN}{RRCNN}{Recursive Residual Convolution Neural Network}
\newacronym{B-DRRN}{B-DRRN}{Deep Recursive Residual Network with Block information}
\newacronym{CPHER}{CPHER}{Coding Prior based High Efficiency Restoration}
\newacronym{MPRN}{MPRN}{Multi-level Progressive Refinement Network}
\newacronym{BDRRN}{BDRRN}{Block Information Constrained Deep Recursive Residual Network}
\newacronym{SEFCNN}{SEFCNN}{Squeeze-and-Excitation Filtering CNN}
\newacronym{GAN}{GAN}{Generative Adversarial Network}
\newacronym{MFRNet}{MFRNet}{Multi-level Feature review Residual dense Network}
\newacronym{IResLB}{IResLB}{Inception and Residual Learning based Block}
\usepackage[hidelinks]{hyperref}
\begin{document}

%\title{Prediction-Aware In-loop Filtering and Post-processing of VVC using CNN}
\title{A CNN-based Prediction-Aware Quality Enhancement Framework for VVC}

% \author{Fatemeh Nasiri, Wassim Hamidouche, Luce Morin, Nicolas Dhollande, and Gildas Cocherel
% %\IEEEmembership{Member, IEEE}
% \thanks{F. Nasiri, W. Hamidouche and L. Morin are with Université Rennes1, INSA Rennes, CNRS, IETR - UMR 6164, 35000 Rennes, France. (emails: \{firstname.lastname\}@insa-rennes.fr).

% F. Nasiri, N. Dhollande and G. Cochorel are with AVIWEST, 35760, Saint-Gr\'{e}goire, France
% (emails: \{ndhollad, gcocherel\}@aviwest.com).

% F. Nasiri, W. Hamidouch and L. Morin are also with IRT b$<>$com, 35510 Cesson-S\'{e}vign\'{e}, France.  

\author{Fatemeh Nasiri, Wassim Hamidouche, Luce Morin, Nicolas Dhollande, and Gildas Cocherel
%\IEEEmembership{Member, IEEE}
\thanks{F. Nasiri, W. Hamidouche and L. Morin are with University of Rennes 1, INSA Rennes, CNRS, IETR - UMR 6164, 35000 Rennes, France. (emails: \{firstname.lastname\}@insa-rennes.fr). 

F. Nasiri, N. Dhollande and G. Cocherel are with AVIWEST, Parc Edonia, Batiment X1, Rue de la Terre de Feu, 35760 Saint-Gr\'{e}goire, France
(emails: \{ndhollande, gcocherel\}@aviwest.com).

F. Nasiri, W. Hamidouche and L. Morin are also with IRT b$<>$com, 35510 Cesson-S\'{e}vign\'{e}, France.

}
}

\maketitle

\begin{abstract}
This paper presents a framework for \gls{cnn}-based quality enhancement task, by taking advantage of coding information in the compressed video signal. The motivation is that normative decisions made by the encoder can significantly impact the type and strength of artifacts in the decoded images. In this paper, the main focus has been put on decisions defining the prediction signal in intra and inter frames. This information has been used in the training phase as well as input to help the process of learning artifacts that are specific to each coding type. Furthermore, to retain a low memory requirement for the proposed method, one model is used for all \glspl{qp} with a \gls{qp}-map, which is also shared between luma and chroma components. In addition to the \gls{pp} approach, the \gls{ilf} codec integration has also been considered, where the characteristics of the \gls{gop} are taken into account to boost the performance. The proposed \gls{cnn}-based \gls{qe} framework has been implemented on top of the \gls{vvc} Test Model (VTM-10). Experiments show that the prediction-aware aspect of the proposed method improves the coding efficiency gain of the default \gls{cnn}-based \gls{qe} method by 1.52\%, in terms of \acrshort{bd-br}, at the same network complexity compared to the default \gls{cnn}-based \gls{qe} filter. 
\end{abstract}

\glsresetall

\section{Introduction}
\IEEEPARstart{V}{ideo} codecs aim at reducing the bit-rate of compressed videos to decrease the traffic pressure on the transmission networks. As this process directly affects the perceived quality of received videos, the importance of retaining high quality displayed video becomes more evident. In particular, the emergence of new video formats, such as immersive 360$^{\circ}$, 8K and \gls{vr}, has pushed more pressure on further bandwidth saving in order to guarantee an acceptable quality. To address this problem, in recent years, besides the great improvements in the domain of transmission network technologies, the development of new video codecs and standards has been initiated. Notably, \gls{vvc} ~\cite{vvc}, AoM Video codecs (AV1 and AV2) \cite{av1} and \gls{evc}~\cite{choi2020overview} are expected to bring a significant improvement in terms of bitrate saving over existing video coding standards such as \gls{hevc}~\cite{sullivan2012overview}. 

Although the new video codecs benefit from more efficient algorithms and tools compared to the previous generation standards, reconstructed videos using these codecs still suffer from compression artifacts, especially at low and very low bitrates. The block-based aspect of the hybrid lossy video coding architecture, shared among all these codecs, is the main source of the blockiness artifact in reconstructed videos. %\hl{sould be associated with QP, it is the only lossy part}. 
To remove this type of artifact, a \gls{dbf} has been used in most of existing codecs \cite{vvc,sullivan2012overview,av1, vvcinloop}. \gls{dbf} applies low pass filters in order to smooth out block borders and correct the discontinuous edges across them. %LM : mention "inloop" (or too early?)
Quantization of transform coefficients rather introduces other types of compression artifacts, such as blurriness and ringing. The larger the quantization step gets, the more visible the blurriness and the ringing become. The quantization step is controlled by \gls{qp}, which varies from 1 to 63 in \gls{vvc}. In low bit-rate video coding, where higher \gls{qp} values are used, the perceived quality is visibly degraded. \gls{sao} and \gls{alf} %LM filters/steps 
are additional filters that are mainly designed to overcome this problem. \Gls{sao} categorizes reconstructed pixels into pre-trained classes and associates to them a set of optimized offsets to be transmitted for texture enhancement. \Gls{alf}, that is applied after \gls{dbf} and \gls{sao} in \gls{vvc}, further improves reconstructed frames. In \gls{alf}, parameters of a set of low pass filters are optimized at the encoder side and transmitted to the decoder.
The common aspect between all these methods is the hand-crafted nature of their algorithms. Although these methods significantly remove undesirable artifacts, the task of enhancing reconstructed videos still has room for further quality improvement. 

The promising advances in the domain of machine learning have recently encouraged the broadcast industry to explore it in the video  compression  domain.  Particularly,  deep  \glspl{cnn} have attracted more attention owing to their significant performance \cite{liu2020deep,ma2019image}. Motivated by \gls{cnn}-based approaches in other image processing tasks, such as \acrfull{sr} and machine vision, several recent studies have been established in the domain of artifact removal from compressed videos. These approaches are categorized into two main groups: \acrfull{pp} and \acrfull{ilf}. The \gls{pp} approach improves reconstructed videos after the decoding step and is considered flexible in terms of implementation, as it is not normatively involved in the encoding and decoding processes. In other words, such a \gls{pp} algorithm serves as an optional step to be used based on the hardware capacity of decoder/receiver device. On the contrary, \gls{ilf} approach involves the normative aspect of encoding and decoding, by generating high quality reconstructed frames to be served as a reference to other frames in the prediction process. %This aspect allows them to offer higher bit-rate saving and higher quality, when applied on a smaller sub-set of frames. 

%Significant improvement, in terms of bit-rate saving and perceived quality, due to the use of \gls{cnn}-based \gls{pp} and \gls{ilf}, has gained them a lot of attention in recent studies, especially for performance improvement of \gls{vvc}. 
In order to reduce artifacts and distortions in reconstructed videos, it is essential to take into account the source and the nature of the artifacts. Most studies have only used reconstructed video and corresponding original video as the ground truth for the training phase of their networks \cite{ma2020mfrnet, JVETN0169}. %While these network are supposedly deals with a complex mapping between the reconstructed signal and the ground truth. 
Except for \gls{qp}, which has a key influence on the distortion level, the use of other coding information is mostly overlooked in the existing studies. To further improve this aspect, in some more advanced works, coding information such as partitioning, prediction and residual information are also used \cite{kang2017multi, he2018enhancing,wang2019integrated}. However, these approaches are mainly applied to intra coded frames.   

%LM : be more precise and thus make a synthesis of the issues that you claim to address. position your study with respect to the classification that you have made for previous work. make explicit that you want to use learning and why.  State what is the difference/similarity of your approach with he previous works.
This paper presents a \gls{cnn}-based framework for quality enhancement of compressed video. The key element of the proposed method is the use of prediction information in intra and inter coded frames. %To this end, two methods namely \acrfull{pacp} and \acrfull{paci} are introduced and integrated in \gls{vvc} as post processing and in-loop filtering, respectively. In both implementations, separate networks are trained for intra and inter coded frames 
%LM : en of stence bellow not convincing: the training does not remove the artifacts. explain what is the reason why you want to train separately 
To this end, a prediction-aware \gls{qe} method is proposed and used as the core module of two codecs integration approaches in \gls{vvc}, corresponding to \gls{pp} and \gls{ilf}. In this method, separate models are trained for intra and inter coded frames to isolate the learning of their specific artifacts. The proposed framework emphasizes on the prediction type as critical coding information and offers frame-level as well as block-level granularity for enhancing the quality of reconstructed video pixels. %The codec integration of the proposed framework has been carried out in the latest version of the VVC Test Model (VTM-10.0), resulting in coding efficiency gain in different coding configurations (\textit{e.g.} All Intra, Random Access \textit{etc.}). 
The main contributions of this paper are summarized as follows:
\begin{enumerate}
    \item Design and implementation of a complete framework for \gls{cnn}-based quality enhancement based on frame-level and block-level prediction types in VTM-10.    
    \item Use of normative prediction decisions for training and testing of both intra and inter coding modes. %For intra coding mode, proposing an approach to take into account the normative decisions made by the encoder regarding spatial texture modelling via intra prediction modes. Likewise, for inter coding mode, exploiting normative decisions of motion modelling to inform the network about temporally correlated content, possibly with higher quality texture.% due to the hierarchical \gls{qp} cascading of temporal layers.
    \item In inter frames, offering block-level granularity for distinguishing between enhancement task of intra blocks, inter blocks and skip blocks, using a block-type mask.
    \item In-loop implementation of the proposed framework, using a normative frame-level signalling to deactivate \gls{cnn}-based enhancement in case of quality degradation. 
    \item Finally, minimizing the memory requirement by sharing the \gls{qe} models between all three colour components in all \gls{qp} values.
    %\item Finally, presenting results of several experiments to analyse the impact of each significant design choice.
\end{enumerate}

The remaining of this article is organized as follows. Section~\ref{sec:relatedwork} introduces the related works and categorizes them based on their contribution and relevance to this work. Section \ref{sec:proposed} presents details of the proposed prediction-aware \gls{qe} method. This method is then used in Section~\ref{sec:codec} as the core \gls{qe} module in two codec integration approaches (\gls{pp} and \acrshort{ilf}). Experiment results are presented in Section~\ref{sec:results}, and finally the paper is concluded in Section \ref{sec:conclusion}.

\section{Related Work}
\label{sec:relatedwork}
In this section, some studies with significant contribution to the \gls{cnn}-based video \gls{qe} task are reviewed. As the proposed framework of this paper particularly focuses on the use of coding information, these studies are categorized and ordered to reflect how much they take into account the nature of compressed video signal and how their method exploits spatial and temporal correlations for the \gls{qe} task. Table~\ref{tab:literature} provides a list of the most relevant papers published in the past few years and summarizes their principle contributions, with a focus on the use of coding information.
%LM end of sentence bellow not clear to me : do you mean "and how the method exploits spatial and temporal correlations"?
%and the fact video compression exploits spatial and temporal correlations.

\begin{table*}[!ht]
\caption{An overview of recently published CNN-based \gls{qe} methods in the literature, with a summary of their contribution as well as the type of coding information they use (e.g. \acrfull{tum} , \acrfull{cum}, \acrfull{resm}),
\acrfull{predm},
\acrfull{mm},
\acrfull{ipmm} and \acrfull{ct}. Moreover, the functionality (Func.) denotes whether a method is integrated in codec as \gls{ilf} or \gls{pp}.}
\centering
\scalebox{1}{
\centering
\renewcommand{\arraystretch}{1.15} 
\begin{tabular}{P{0.09\textwidth} P{0.12\textwidth} P{0.06\textwidth} P{0.09\textwidth} P{0.02\textwidth} C{0.48\textwidth}}

\hline
\hline
% p{0.1\textwidth}p{0.8\textwidth}
%& & & & \\
Method	&	Published		&Training dataset&Coding information	& QE Func	& \multicolumn{1}{c}{Summary of contribution} \\	 \thickhline
IFCNN	&	IVMSP	16	\cite{park2016cnn}	&	CTC	&	QP	&	\acrshort{ilf}	&	Applied after DB instead of SAO. 3 layers CNN with residual learning.	\\	\hline
STResNet	&	VCIP	17	\cite{jia2017spatial}	&	AVS2	&	QP	&	\acrshort{ilf}	&	 After SAO, uses previous reconstructed block. 4 CNN layers with residual learning  	\\	\hline
DCAD	&	DCC	17	\cite{wang2017novel}	&	-	&	QP	&	\acrshort{pp}	&	 A 10 layers CNN network with residual learning. 	\\	\hline
DSCNN	&	ICME	17	\cite{yang2017decoder}	&	BSDS500	&	QP	&	\acrshort{pp}	&	Scalable network with separate branches for inter and intra frames	\\	\hline
MMS-net	&	ICIP	17	\cite{kang2017multi}	&	Xiph.org	&	QP \acrshort{tum}	&	\acrshort{pp}	&	Replaces all HEVC loop filters in intra coded frames. Scalable training. 	\\	\hline
VRCNN	&	MMM	17	\cite{dai2017convolutional}	&	-	&	QP	&	\acrshort{pp}	&	Replaces HEVC in-loop filters in intra mode, variable CNN filter sizes. 	\\	\hline
MSDD	&	DCC	18	\cite{wang2018multi}	&	Youtube	&	QP	&	\acrshort{pp}	&	Multi-frame input (next and previous frames) with multi-scale training	\\	\hline
-	    &	ICIP	18	\cite{he2018enhancing}	&	-	&	QP \acrshort{pm} \acrshort{mm}	&	\acrshort{pp}	&	Mean and partitioning mask with reconstructed frames are fed to a residual-based net.	\\	\hline
QECNN	&	  IEEE-TCSVT	18	\cite{yang2018enhancing}	&	BSDS500	&	QP	&	\acrshort{pp}	&	Two networks with different filter sizes for inter and intra frames, time constrained QE	\\	\hline
-	    &	IEEE access	18	\cite{soh2018reduction}	&	JCT-VC	&	QP	&	\acrshort{pp}	&	Temporally adjacent similar patches are also fed to an inception-based net.  	\\	\hline
MFQE	&	CVPR	18	\cite{yang2018multi}	&	JCT-VC	&	QP	&	\acrshort{pp}	&	Current and motion compensated frames of high quality adjacent frames are fed to net.  	\\	\hline
CNNF	&	ICIP	18	\cite{song2018practical}	& DIV2K	&	QP	&	\acrshort{ilf}	&	QP and reconstructed frame are fed to network, replacing SAO and DBF. 	\\	\hline
RHCNN	&	IEEE-TIP	18	\cite{zhang2018residual}	&	CTC	&	QP	&	\acrshort{ilf}	&	 QP-specific training of a network based on several residual highway units	\\	\hline
FECNN	&	ICIP	18	\cite{li2018deep}	&	BSDS500	&	QP	&	\acrshort{pp}	&	A residual based network with two skip connections proposed for intra frames.	\\	\hline
Residual-VRN	&	BigMM	18	\cite{ma2018residual}	&	MSCOCO	&	QP \acrshort{resm}  \acrshort{predm}	&	\acrshort{pp}	&	Prediction and quantized residual frame are fed to a residual based network as input.	\\	\hline
MGANet	&	arXiv 	18	\cite{meng2018mganet}	&	Derf	&	QP \acrshort{tum}	&	\acrshort{pp}	&	\acrshort{tum} is also fed to a multi-scale net. which exploits output of a temporal encoder. 	\\	\hline
ADCNN	&	IEEE access	19	\cite{wang2019attention}	&	DIV2K	&	QP \acrshort{tum}	&	\acrshort{ilf}	&	Network composed of attentions blocks, using also QP and TU map.	\\	\hline
MM-CU-PRN	&	ICIP	19	\cite{wang2019partition}	&	DIV2K	&	QP \acrshort{cum}	&	\acrshort{pp}	&	Based on Progressive Rethinking Block which multi-scale CU maps are also used.  	\\	\hline
SDTS	&	ICIP	19	\cite{meng2019enhancing}	&	CTC	&	QP	&	\acrshort{pp}	&	Multi frame QE scheme, using motion compensated frames and an improved network.	\\	\hline
VRCNN-BN	&	IEEE access	19	\cite{zhao2019cnn}	&	-	&	QP	&	\acrshort{pp}	&	Adds further non-linearity to VRCNN by adding batch normalization and Relu layers. 	\\	\hline
MIF	&	IEEE-TIP	19	\cite{li2019deep}	&	HIF	&	QP \acrshort{cum} \acrshort{tum}	&	\acrshort{ilf}	&	Selects high quality references to the current frame and exploits them in the QE.	\\	\hline
-	&	APSIPA ASC	19	\cite{wang2019integrated}	&	DIV2K	&	QP \acrshort{cum} \acrshort{tum}	&	\acrshort{pp}	&	 A network based on residual learning, exploiting TU and CU maps.	\\	\hline
MRRN	&	IEEE SPL	19	\cite{yu2019quality}	&	BSDS500	&	QP	&	\acrshort{pp}	&	Adopts a multi-reconstruction recurrent residual network for PP-QE task. 	\\	\hline
RRCNN	&	  IEEE-TCSVT	19	\cite{zhang2019recursive}	&		&	QP	&	\acrshort{ilf}	&	Intra , Recursive structure and Residual units with local skip connections	\\	\hline
B-DRRN	&	PCS	19	\cite{hoang2019b}	&	-	&	QP \acrshort{cum} \acrshort{mm}	&	\acrshort{pp}	&	Network based on recursive residual learning, exploiting mean and boundary mask.	\\	\hline
CPHER	&	ICIP	19	\cite{feng2019coding}	&	Vimeo	&	QP \acrshort{predm}  	&	\acrshort{pp}	&	Network based on residual blocks, exploiting unfiltered frame and prediction.	\\	\hline
WARN	&	ICIP	19	\cite{chen2019av1}	&	DIV2K 	&	QP	&	\acrshort{ilf}	&	A wide activation residual network for ILF of AV1 codec. 	\\	\hline
ACRN	&	ICGIP	19	\cite{bei2020cu}	&	DIV2K 	&	QP	&	\acrshort{ilf}	&	Asymmetric residual network as ILF in AV1, with a more complex net. for higher QPs. 	\\	\hline
QG-ConvLSTM	&	ICME	19	\cite{yang2019quality}	&	-	&	QP	&	\acrshort{pp}	&	Spatial features of distorted frame are combined with neighboring frames' features.  	\\	\hline
-	&	IEEE-TIP	19	\cite{lu2019deep}	&	Vimeo	&	QP	&	\acrshort{pp}	&	Based on Kalman filters, using temporal information restored from previous frames.  	\\	\hline
SimNet	&	PCS	19	\cite{ding2019cnn}	&	 DIV2K	&	QP	&	\acrshort{pp}	&	Depth of network is varied based on the distortion level.	\\	\hline
LMVE	&	ICIP	19	\cite{tong2019learning}	&	-	&	QP	&	\acrshort{pp}	&	Single and multi frame QE net. proposed, using FlowNet to generate high quality MC. 	\\	\hline
-	&	CVPR	19	\cite{lu2019learned2}	&	DIV2K	&	QP	&	\acrshort{pp}	&	 Residual block based network which receives different scales of input frame. 	\\	\hline
SEFCNN	&	  IEEE-TCSVT	19	\cite{ding2019switchable}	&	DIV2K	&	QP \acrshort{ct}	&	\acrshort{ilf}	&	Optional ILF with adaptive net. selection for different CT and distortion levels.	\\	\hline
DIANet	&	PCS	19	\cite{xu2019dense}	&		&	QP	&	\acrshort{ilf}	&	 Dense inception net. with different attention blocks, separating inter/intra frames.	\\	\hline
-	&	IEEE-TIP	19	\cite{jia2019content}	&	BSDS500	&	QP	&	\acrshort{ilf}	&	 Content-aware ILF with adaptive network selection depending on CTU content	\\	\hline
MFRNet	&	arXiv 	20	\cite{ma2020mfrnet}	&	BVI-DVC	&	QP	&	\acrshort{ilf}	&	An architecture based on multi-level dense residual blocks with feature review. \\	\hline
EDCNN	&	IEEE-TIP	20	\cite{pan2020efficient}	&	CTC	&	QP	&	\acrshort{ilf}	&	Network with enhanced residual blocks with weight normalization.	\\	\hline
BSTN	&	MIPR	20	\cite{meng2020bstn}	&	Derf SJTU 	&	\acrshort{tum} \acrshort{mm}	&	\acrshort{pp}	&	MC frames along with distorted frame and additional coding information are fed to net.	\\	\hline
FGTSN	&	DCC	20	\cite{meng2020flow}	&	-	&	QP	&	\acrshort{pp}	&	Flow-guided multi-scale net. using motion field extracted from neighboring frames.	\\	\hline
- &	IEEE access	20	\cite{chen2020neural}	&	JCT-VC	&	QP	&	\acrshort{pp}	&	 Sparse coding based reconstruction frame fed to net. with MC and distorted frames.	\\	\hline
-	&	ACM 	20	\cite{lam2020efficient}	&	CLIC	&	QP	&	\acrshort{pp}	&	Fine-tuned QE network transmitting modified weights via bitstream. 	\\	\hline
FQE-CNN	&	  IEEE-TCSVT	20	\cite{huang2020frame}	& CLIC	&	QP \acrshort{ipmm}	&	\acrshort{pp}	&	Image size patches used for training, using intra modes map. 	\\	\hline
-	&	ICME	20	\cite{zhang2020enhancing}	&	BVI 	&	QP	&	\acrshort{pp}	&	Post-processing for VVC encoded frames with network based on residual blocks.	\\	\hline
RBQE	&	arXiv 	20	\cite{xing2020early}	&	RAISE	&	QP	&	\acrshort{pp}	&	Blind QE with an easy-to-hard paradigm, based on dynamic neural net.	\\	\hline
IFN/PQEN-ND	&	NC 	20	\cite{sun2020quality}	&	SJTU	&	QP	&	\acrshort{ilf}	&	Noise characteristic extracted from frames for enhancing intra and inter frames.	\\	\hline
QEVC	&	ICCCS	20	\cite{li2020qevc}	&	CDVL,REDS 	&	QP 	&	\acrshort{ilf}	&	Depending on motion, a selector network selects different networks for QE.\\	\hline
MSGDN	&	CVPRW	20	\cite{li2020multi}	&	COCO2014	&	QP	&	\acrshort{pp}	&	A multi-scale grouped dense network as a post-processing of VVC intra coding 	\\	\hline
PMVE	&	IEEE-TC	20	\cite{ding2020biprediction}	&	-	&	QP	&	\acrshort{pp}	&	Frames are enhanced by contributing prediction info. from neighboring HQ frames.\\	\hline
MWGAN	&	arXiv 	20	\cite{wang2020multi}	&	-	&	QP	&	\acrshort{pp}	&	GAN multi-frame wavelet-based net., recovering high frequency sub-bands.	\\	\hline
STEFCNN	&	DCC	20	\cite{xu2020spatial}	&	-	&	QP	&	\acrshort{pp}	& Multi-frame QE method with a dense residual block based pre-denoising stage 	\\	\hline
MPRNET	&	NC 	20	\cite{jin2020post}	&	BSD500	&	QP	&	\acrshort{pp}	& GAN multi-level progressive refinement, replacing the DBF and SAO in HEVC	\\	\hline
RRDB	&	UCET	20	\cite{wang2020visual}	&	BSD500	&	QP	&	\acrshort{pp}	&	A GAN-based network for QE of Intra coded frames of HEVC	\\	\hline

\hline
\hline
\end{tabular}
\label{tab:literature}

}
\end{table*}

\subsection{Single-Frame Quality Enhancement} 
Most \gls{cnn}-based \gls{qe} methods enhance the quality of video in a frame-by-frame manner, where each frame is enhanced independently. These methods exploit spatial information of texture pixels of individual frames in order to enhance their quality and remove their artifacts. One of the early works in this category, proposed in \cite{park2016cnn}, uses a network with three convolutional layers to learn the residual information. This method is implemented as \gls{ilf} and replaces \gls{sao} filter of \gls{hevc}. Another method called \gls{DCAD}, deploys a relatively deeper network with ten layers to be used as a \gls{pp} filter after decoding \cite{wang2017novel}. Inspired by the diversity of block sizes in \gls{hevc}, an \gls{ilf} named \gls{VRCNN} proposes a network with different filter sizes to replace both \gls{sao} and \gls{dbf} filters of \gls{hevc} intra coding \cite{dai2017convolutional}. The method presented in \cite{zhao2019cnn} enhances the performance of \gls{VRCNN} by introducing more non-linearity to the VRCNN network. The added ReLU \cite{nair2010rectified} and batch normalization \cite{ioffe2015batch} layers in this method improve its performance, compared to \gls{VRCNN}.

In another work, presented in \cite{yang2018enhancing}, two different networks are trained for intra and inter frames. The intra network is a sub-net of the inter network which helps the method to capture artifacts of intra coded blocks in P and B frames more efficiently. Furthermore, in \cite{yang2018enhancing}, the complexity of the \gls{qe} filter is controlled by comparison of a \gls{mse}-based distortion metric in the \gls{ctu} level at the encoder side. The \gls{RHCNN} network, presented in \cite{zhang2018residual}, is composed of several cascaded residual blocks, each of which having two convolutional layers, followed by a ReLU activation function. In \gls{RHCNN}, inter and intra coded frames are also enhanced with dedicated networks. In \cite{pan2020efficient}, residual blocks in the network are enhanced by splitting the input frame and processing each part with a different \gls{cnn} branches. The output of each branch is then concatenated and fed to the next block. As another contribution, a weighted normalization scheme is used instead of batch normalization which also improves the training process.

More recently, \gls{MM-CU-PRN} loop filter has been introduced  \cite{wang2019partition} which uses progressive rethinking block and additional skip connections between blocks, helping the network to keep a longer memory during the training and be able to use low level features in deeper layers. \gls{MM-CU-PRN} is placed between \gls{dbf} and \gls{sao} filters and benefits from coding information by using multi-scale mean value of \glspl{cu}. The network presented in \cite{ma2020mfrnet}, \gls{MFRNet}, deploys similar network architecture as in \gls{MM-CU-PRN}. However, \gls{MFRNet} utilizes multi-level residual learning while reviewing (reusing) high dimensional features in the input of each residual block which leads to a network with better performance compared to existing networks.

\gls{MRRN} is a method based on recursive learning and is implemented as \gls{pp} filter for decoded frames of \gls{hevc}~\cite{yu2019quality}. In recursive learning, the same layers are repeatedly applied which reduces the probability of over-fitting during the training. Likewise, in \cite{zhang2019recursive}, another recursive residual network is proposed as \gls{ilf} for intra coded frames. The proposed network in \cite{zhang2019recursive} is applied on reconstructed frames before the \gls{dbf} and \gls{sao} filters. \gls{BDRRN} is another method based on recursive residual learning in which a block-based mean-mask, as well as the boundary-mask, are used as input to network \cite{hoang2019b}. 

Furthermore, in some works, the focus is put on strategies for enhancing the quality of video frames. In \cite{xing2020early}, a blind quality enhancement approach is proposed where frames with different distortion levels are processed differently. An ``easy-to-hard'' \gls{qe} strategy is used to determine which level of the CNN-based filtering shall be applied on a given frame. In the case that the quality is already satisfying based on a blind quality assessment metric, the \gls{qe} process stops, otherwise, it continues. In \gls{SEFCNN} \cite{ding2019switchable}, an adaptive \gls{ilf} is also proposed in which networks with various complexity levels are trained for different \glspl{qp}.  

%In \acrfull{MPRN} \cite{jin2020post}, a GAN-based post-processing filter for intra coded frames is presented to be used instead of SAO and DBF. The generator network utilizes a progressive refinement strategy to generate enhanced frames.

\subsection{Multi-frame Quality Enhancement}
Multiframe \gls{qe} methods process a set of consecutive frames as input. The basic idea behind this category of methods is to remove the compression artifacts while considering the temporal correlation of video content. Moreover, the quality is propagated from frames encoded at high quality to adjacent frames encoded at lower quality. 

One of the earliest implementations exploits temporal information simply by adding previously decoded frames to the input of the network, along with current frame~\cite{jia2017spatial}. In another method, \glspl{pqf} are detected with an SVM-based classifier \cite{yang2018multi}. Using a network named \gls{mc} sub-net, the MC frames of previous and next \glspl{pqf} are generated. The three frames are then fed to another network, called \gls{qe} sub-net, in order to enhance the quality of the current frame, while the selected \glspl{pqf} are kept to be enhanced by another dedicated network. An improved version of this method has been introduced in~\cite{guan2019mfqe,meng2019enhancing}, where the high quality frame detection, as well as the \gls{qe} network itself, are improved. 

In multiframe \gls{qe} methods, finding the best similar frames to the current frame for the task of motion compensation is important. In \cite{li2019deep}, similar reference frames with higher quality than the current frame are detected with a dedicated network. Then they are used to generate the \gls{mc} frames with respect to the content of the current frame. This frame with computed motions is then used as input to the \gls{qe} network along with the reconstructed frame. In another research, a flow-guided network is proposed, where the motion field is extracted from previous and next frames using FlowNet~\cite{ilg2017flownet,meng2020flow}. Once the motion compensation is completed, a multi-scale network is applied to extract spatial and temporal features from the input. Following the same principle, motion compensated frames of adjacent frames are fed to network in \cite{meng2020bstn}. A ConvLSTM-based network is then used to implicitly discover frame variations over time between the compensated adjacent frames and the current frame. Moreover, in order to capture the texture distortion in compressed frames, the \gls{tu} mean map is also fed to the network. In \cite{chen2020neural}, in addition to the motion compensated frame, a sparse coding based reconstruction frame is also fed to the network as input. The purpose of using sparse coding prediction is to simplify the process of texture learning by the network. Similarly, in \cite{soh2018reduction}, most similar patches in previous and next frames are extracted and fed to a network with three branches of stacked convolutional layers. The branches are then concatenated to reconstruct the final patch. 

Inspired by the multi-frame \gls{qe} methods, a bi-prediction approach is proposed in \cite{ding2020biprediction}. In this work, instead of computing the motion field, a prediction of the current low quality frame is generated from neighbouring high quality frames. Then the predicted frame and reconstructed frame are fed to the \gls{qe} network. In \cite{wang2020multi}, a \gls{GAN}-based multi-frame method is presented in which adjacent frames and current frame are fed to a \gls{GAN}, which is itself composed of two parts: one for the motion compensation and the other for quality enhancement. Wavelet sub-bands of motion compensated frames and reconstructed frames are fed to the second part of the network as input. For evaluation of generated content, a wavelet-based discriminator that extracts features in the wavelet domain at several levels (\textit{i.e.} sub-bands) is proposed.

%\hl{this sentence is not clear how we do evaluation with wavelet discriminator} -- \hl{\textit{is it ok now?}}. 

% our contribution:
In summary, multi-frame solutions adopt different motion compensation methods mostly based on block-matching or \gls{cnn}-based approaches. However, they all overlook the fact that the actual motion modelling is performed based on a normative process which takes into account complex factors such as bitrate restriction or internal state of the encoder modules. In other words, one might consider the normative motion information available in the bitstream more useful than texture-based heuristic motion modelling. Furthermore, this signal is already available both at the encoder and decoder sides and can easily be used as side information for inference in the \gls{qe} networks of \gls{pp} and \gls{ilf} methods.

\subsection{Methods based on coding information}
In the literature, there are diverse levels of involving coding information in the \gls{cnn}-training of the \gls{qe} task. Here, we categorize these methods from the most basic coding information to the most advanced ones.

\subsubsection{\acrfull{qp}} A common basic coding information and one of the most useful ones is \gls{qp}. Most of methods somehow involve the applied \gls{qp} of the encoded signal in the training and the inference phases. There are mainly two approaches to use \gls{qp} in \gls{cnn}-based \gls{qe}: 
\begin{itemize}
    \item \gls{qp}-specific training: dedicating one model for each \gls{qp} or a range of \glspl{qp}~\cite{ma2020mfrnet,nasiri2020prediction}.
    \item \gls{qp}-map training: providing \gls{qp} as an input to the network \cite{wang2019attention, wang2019integrated, zhang2019recursive, lam2020efficient, ding2020biprediction}.
\end{itemize}
Each approach has benefits and drawbacks. In the \gls{qp}-specific training methods, the performance is usually higher as the artifacts of each \gls{qp} have been particularly observed by their dedicated network during the training. However, they usually require storing several trained models at the decoder-side which is not hardware-friendly. On contrary, \gls{qp}-map methods are usually lighter to implement, especially when the \gls{qp} value varies in finer granularity such as frame-level or block-level.

\subsubsection{Partitioning} Another common coding information is block partitioning and boundary information. Depending on the flexibility of the codec under study (\textit{e.g.} \gls{hevc}, AV1, \gls{vvc}), this aspect is used differently in the literature. The simplest form of partitioning information is the boundary mask \cite{he2018enhancing}. More sophisticated methods, especially \gls{hevc}-based ones, differentiate between \acrfullpl{cu}, \acrfullpl{pu} and \acrfullpl{tu} boundaries
~\cite{kang2017multi, he2018enhancing, wang2019attention, li2019deep, wang2019integrated, hoang2019b, meng2020bstn}. 

\subsubsection{Prediction information} Spatial and temporal prediction information has also been used for enhancement of coded videos. 
%In intra coding mode, the simplest prediction representation is the mean-mask of intra blocks \cite{he2018enhancing}. Other methods use actual intra prediction signal associated to intra blocks. 
In \cite{he2018enhancing}, coding information such as the partitioning map as well as a mean-mask have been used as input to their proposed \gls{cnn}-based \gls{qe} network. The mean-mask is computed based on average reconstructed pixel values in each partition. %In this method, a network with several \gls{cnn}-based residual units is used which takes two signals as input: reconstructed frame, mean-mask. 
In another work, presented in \cite{feng2019coding}, a \gls{qe} network is proposed in which the unfiltered frame and prediction frame are used along with the reconstructed frame as the input of the network. Finally, a three-level network composed of \acrfull{IResLB} units is proposed. The \gls{IResLB} units have three branches, each one having one to two convolutional layers. The intra mode map is then fed to the network to enhance the intra coded frames~\cite{huang2020frame}.

Regarding inter coding mode, an \gls{ilf} with a selector network has been proposed in \cite{li2020qevc}. In this method, the selector 
%\hl{selection or selector ?}  % it is selector in the original paper
network determines the motion complexity of a set of selected \glspl{cu} and then decides whether to increase or decrease the \gls{qp} value of CU and also which network (large or small scale network) to be used. 

\subsubsection{Residual information} Finally, the residual information has also been used as an additional input information for the task of encoded video enhancement. In \cite{ma2018residual}, the coded residual information and prediction information are fed to a network to enhance intra coded frames in \gls{hevc}. The proposed network uses \gls{dc}-ReLU activation function in the first residual layer. The loss function in this work is a combination of \gls{msssim}, L1 and L2 functions. In another work, presented in \cite{lu2018deep}, the \gls{qe} task is modelled as a Kalman filtering procedure and enhance the quality through a deep Kalman network. To further improve the performance of the network, it uses prediction residuals as prior information. 
%\hl{here you should put the transition to the next section what is new in your solution with respect what you described}

To best of our knowledge, it is the first \gls{qe} framework in which the spatial and temporal prediction information in frame and block levels are used for compressed video. In the following sections, the details of proposed framework are explained and integration in the \gls{vvc} codec at both \gls{pp} and \gls{ilf} is presented.  

% ----------------------------------------------
% ----------------------------------------------
% ----------------------------------------------

%\section{Proposed Method}
\section{Proposed Quality Enhancement Neural Networks}
\label{sec:proposed}
In this section, fundamental elements of the proposed prediction-aware \gls{qe} method are described. A common network architecture is adopted that takes into account prediction information associated with reconstructed image. This network is then trained separately for intra and inter images, and applied at the frame-level and block-level to both luma and chroma components, in order to enhance their content based on local coding types.

%\hl{Luce suggested to mention somewhere and clarify the vocabulary about ``residual blocks'' when they are used in the context of video compression or network architecture. Is it a good place here? Or should I put it in the footnote?}
%-------------------------------------------------------------------------------------------------
%-------------------------------------------------------------------------------------------------

\subsection{Network architecture}
\begin{figure*}[ht]
    \centering
    \includegraphics[scale=.25]{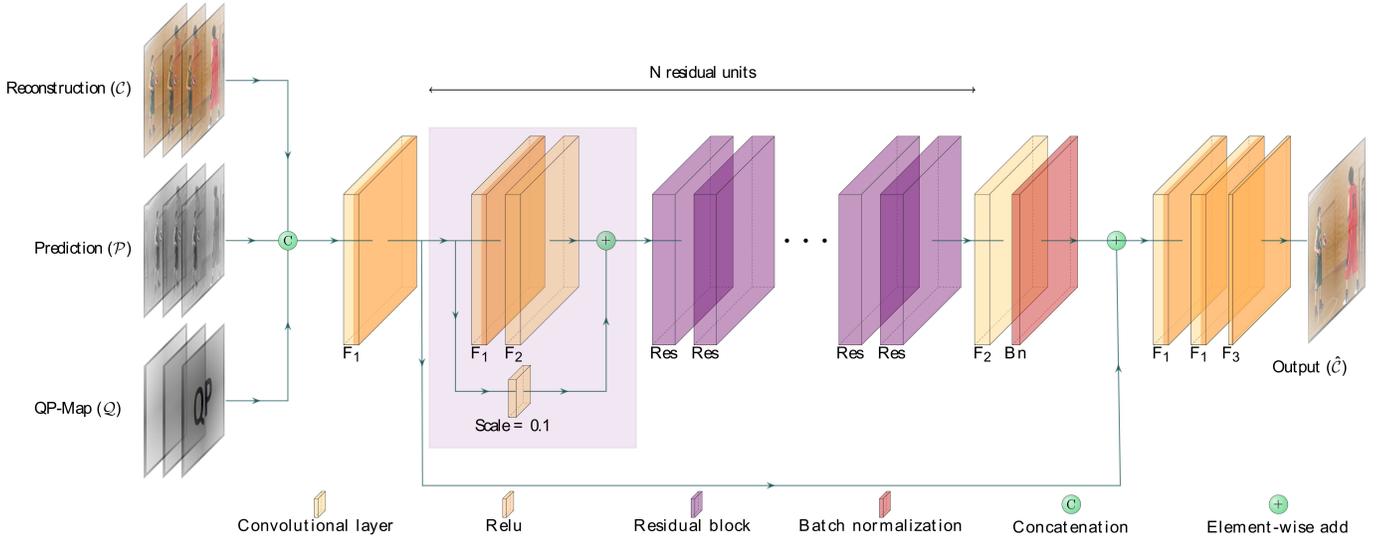}
    \caption{Network architecture of the proposed method using the prediction, QP-map and reconstruction signal as the input.}
    \label{arch}
\end{figure*}

Recently there have been numerous studies on \gls{cnn}-based architectures, improving their performance and complexity. In the literature, the residual blocks~\footnote{In the remaining of this paper, the term ``residual'' is occasionally used interchangeably in the context of video compression residual signal as well as neural network residual layer.} have been widely used for super resolution and quality enhancement tasks which results in better enhancement and detail retrieval~\cite{lim2017enhanced,nasiri2020study}. Inspired by those works, the network architecture of this paper is based on residual blocks, combined with \gls{cnn} layers, as shown in Fig~\ref{arch}. 
The first convolutional layer of the adopted network receives a single reconstructed signal as well as two associated coding information. In particular, a \gls{qp} map and the prediction signal, both with the same size as the reconstructed signal, are concatenated with the reconstructed image as coding information. After one convolutional layer, $N$ identical residual blocks, each composed of two convolutional layers and one ReLU activation layer in between, are used. The convolutional layers in the residual blocks have the same size as the feature maps and kernel size of the first convolutional layer. In order to normalize the feature maps, a convolutional layer with batch normalization is applied after the residual blocks. A skip connection between the input of the first and the last residual block is then used. Finally, three more convolutional layers after the residual blocks are used for reconstructing the enhanced reconstructed signal. Worthy of mention, the input size is arbitrary and not limited to one full frame. As will be explained later, depending on the granularity of the \gls{qe} task, the inference might be applied in the frame-level or block-level.

Given $\mathcal{I}$ as the concatenation of the input signals, the process of producing the enhanced reconstructed signal $\mathcal{\hat{C}}$, by the proposed \gls{cnn}-based \gls{qe} method is summarized as:  %\hl{in Fig. 1 you dont have correct notations}:
\begin{equation}
\label{eq:verbose}
\mathcal{\hat{C}}= F_3^1(F_1^2(Bn^1(F_2^1(Res^{N}(F_1^1(\mathcal{I})))) + F_1^1(\mathcal{I}))),
\end{equation}
where $F_1(.)$ and $F_2(.)$ are $3\times 3 \times 256$ convolutional layers, with and without the ReLU activation layer, respectively. Moreover, $F_3(.)$ is a $3\times 3 \times 1$ convolutional layer with the ReLU activation layer. The superscript of each function indicates the number of times they are repeated sequentially in the network architecture. Finally, $Res$ and $Bn$ are the residual block and batch normalization layer, respectively.

Two different sets of inputs are used for the reference and the proposed methods. In the reference method, referred to as the prediction-unaware method in the rest of this paper, only the reconstructed $\mathcal{C}$ and the \gls{qp} map $\mathcal{Q}$ are used, hence:
\begin{equation}
\label{eq:predunaware}
\mathcal{I}_{ref}=\mathcal{C \oplus Q},
\end{equation}
where $\oplus$ is the concatenation operator. Correspondingly, in the proposed method, referred to as the prediction-aware \gls{qe} method, the prediction signal $\mathcal{P}$ is also used:

\begin{equation}
\label{eq:predaware}
\mathcal{I}_{pro}=\mathcal{P\oplus C \oplus Q}.    
\end{equation}

The normalized \gls{qp}-map ($\mathcal{Q}$) for a frame (or a block) with the width and height of $W$ and $H$, respectively, is calculated as:
\begin{equation}
    \mathcal{Q}_{i,j} = \frac{q_{i,j}}{q_{max}}, 
\end{equation}
where $q_{i,j}$ is the \gls{qp} value of the block that contains the pixel at coordinates $(i,j)$, with $0\leq i < W; 0 \leq j < H$, and $q_{max}$ is the maximum \gls{qp} value (\textit{e.g.} 63 in \gls{vvc}).

Regardless of the \gls{qe} method (\textit{ref} or \textit{pro}), the \gls{qe} task can be summarized as: 
\begin{equation}
\label{eq:summary}
%\mathcal{\hat{O}}=f_{QE}(\mathcal{C}, \mathcal{P}, \mathcal{Q}; \theta_{QE}).
\mathcal{\hat{C}}=f_{QE}(\mathcal{I}; \theta_{QE}),
\end{equation}
where $\theta_{QE}$ is the set of parameters in the network architecture of Eq.~\eqref{eq:verbose}. This parameter set is optimized in the training phase, using the $L_1$ norm as the loss function, computed with respect to the original signal $\mathcal{O}$:
\begin{equation}
    L_1(\mathcal{O}-\mathcal{\hat{C}})=|\mathcal{O}-\mathcal{\hat{C}}|.
\end{equation}

%-------------------------------------------------------------------------------------------------
%-------------------------------------------------------------------------------------------------

\subsection{Prediction-aware QE}
Even though the network architecture is common in the proposed method, signal $\mathcal{P}$ is computed differently for intra and inter frames. Hence, different models are trained for each coding type in order to more efficiently learn their distortion patterns and based on their characteristics and functionality.

\subsubsection{Intra coded frames}

\begin{figure*}

    \centering
    \begin{center}
    \begin{tabular}{ ccccc  }

    {Prediction} 
	& 
	{Reconstruction} 
	& 
	{Loss} 
	& 
	&
	{Original luma content of the selected block $\mathcal{O}^k_i$}
\\ 
    {$\mathcal{P}^k_i$} 
	& 
	{$\mathcal{C}^k_i$} 
	& 
	{$\mathcal{O}^k_i$ - $\mathcal{C}^k_i$} 
	& 
	&	
	\multirow{14}{*}{\includegraphics[scale=0.75]{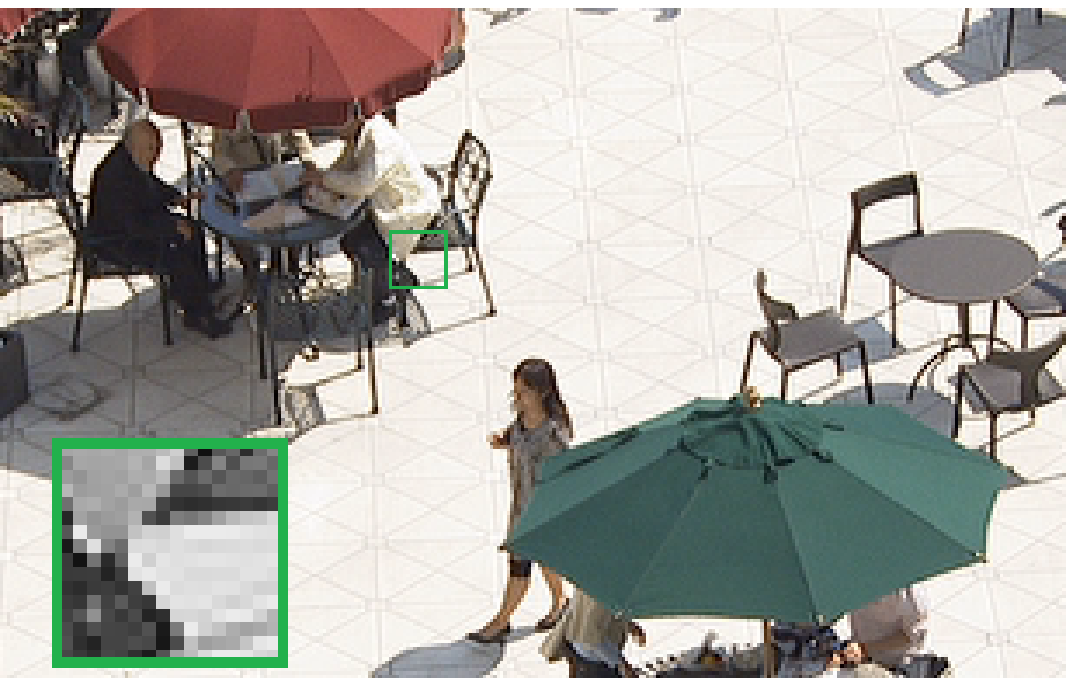}}
\\ 
    & & & &
\\    
    \multirow{4}{*}{\includegraphics[scale=0.8]{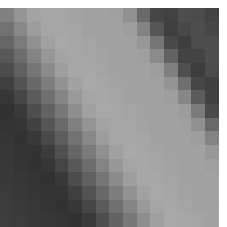} }
    & 
    \multirow{4}{*}{\includegraphics[scale=0.8]{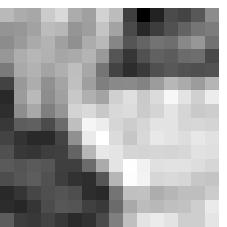} }
    & 
    \multirow{4}{*}{\includegraphics[scale=0.8]{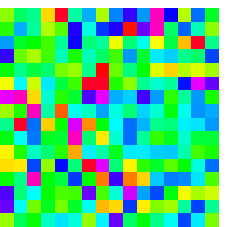}}
    & 
    {IPM 38}    
	&
\\
    & & & {R$_{38}$: 182} &
\\ 
    & & & {D$_{38}$: 22970} &
\\
    & & & {J$_{38}$: 77803} &
\\ 
    & & & &
\\ 
    & & & &
\\
    \multirow{4}{*}{\includegraphics[scale=0.8]{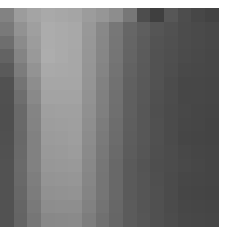} }
    & 
    \multirow{4}{*}{\includegraphics[scale=0.8]{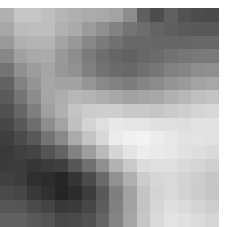} }
    & 
    \multirow{4}{*}{\includegraphics[scale=0.8]{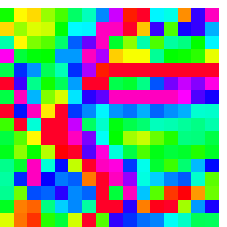}}
    &
    {IPM 50}    
	&
\\
    & & & {R$_{50}$: 112} &
\\ 
    & & & {D$_{50}$: 43518} &
\\
    & & & {J$_{50}$: 77390} &
\\ 
    & & & &
\\ 
    & & & &

\end{tabular}
    \end{center}
    \caption{An example of how two IPMs with similar R-D cost can result in different compression artifacts. The tested 16 $\times$ 16 block, \textit{k}, is coded with IPMs 38 and 50 in \gls{qp} 40 with $\lambda$=301.}
    \label{intraprediction}
\end{figure*}

Intra coding is based on exploiting the spatial redundancies existing in frame textures. In \gls{vvc}, a set of 67 \gls{ipm}, representing 65 angular \gls{ipm}s, plus \gls{dc} and planar modes are used for modelling texture of blocks. The selection of an \gls{ipm} for a block is performed by optimizing the rate-distortion (R-D) cost, denoted as $J_i$:    

\begin{equation}
    J_i = D_i+\lambda \, R_i \text{\hspace{15pt}} i=1, \dots, 67,
    \label{RD}
\end{equation}

where $D_i$ and $R_i$ are the distortion and the rate of using the $i^{th}$ mode, respectively. The Lagrangian multiplier $\lambda$ is computed based on the \gls{qp} which determines the relative importance of the rate and the distortion during the decision making process. In lower bitrates (higher \gls{qp} values), the value of $\lambda$ is higher, meaning that minimization of the rate is relatively more important than minimization of the distortion. Similarly, the opposite principle is applicable to higher bitrates (lower \gls{qp} values).

The best \gls{ipm}, minimizing the R-D cost of a block, is not necessarily the \gls{ipm} that represents the block texture most accurately \cite{nasiri2020prediction}. An example of such a situation is presented in Fig.~\ref{intraprediction}. In this figure, a $16\times16$ block, \textit{k}, is selected and the prediction ($\mathcal{P}_i^k$) and reconstruction ($\mathcal{C}_i^k$) blocks corresponding to its two best \glspl{ipm} in terms of R-D cost are shown. Precisely, these two best \glspl{ipm} are angular modes 38 and 50. As can be seen, despite their similar R-D costs, these two \glspl{ipm} result in very different reconstructed signals, with different types of compression loss patterns. On one hand, \gls{ipm} 38 is able to model the block content more accurately (i.e. smaller distortion $D_{38}$) at the cost of a higher \gls{ipm}/residual signaling rate (i.e. $R_{38}$). On the other hand, \gls{ipm} 50 provides a less accurate texture modeling (i.e. high distortion $D_{50}$) with a smaller \gls{ipm}/residual signaling rate (i.e. $R_{50}$). As a result, these two \glspl{ipm} result in very different types of artifacts for a given block, as can be seen by comparing the corresponding reconstruction blocks (i.e. $\mathcal{C}_{38}^k$ and $\mathcal{C}_{50}^k$). This behavior is due to two different R-D trade-offs of the selected modes.

The above example proves that the task of \gls{qe} for a block, frame or an entire sequence could be significantly impacted by different choices of coding modes (e.g. \gls{ipm}) determined by the encoder. This assumption is the main motivation in our work to use the intra prediction information for the training of the quality enhancement networks.

The example in Fig. \ref{intraprediction} proves that encoder decisions can have major impact on the \gls{qe} task and its performance. Particularly, for intra blocks, we assume that the selected \gls{ipm} for a block carries important information and shall be included in the training of the proposed \gls{cnn}-based \gls{qe} method~\cite{nasiri2020prediction}. Therefore, an intra prediction frame is constructed by concatenating the intra prediction signals in the block-level. This signal is then used as input to the network.

%-------------------------------------------------------------------------------------------------
\subsubsection{Inter coded frames}
\begin{figure}
    \centering
    \input{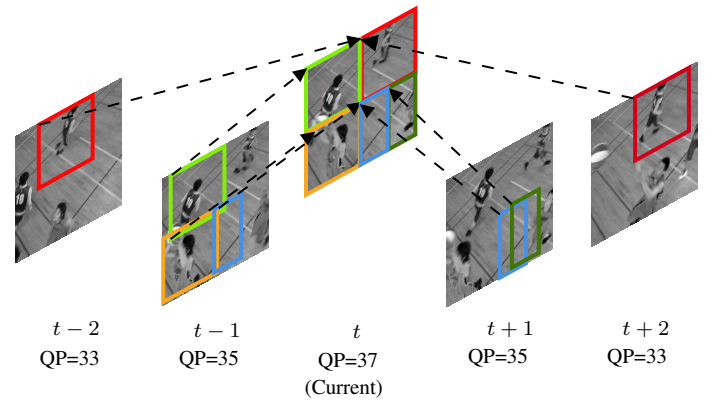}
    \caption{An example of \gls{qp} cascading in the hierarchical \gls{gop} structure, providing higher quality motion compensated blocks at frame $t$ from past ($t-1$ and $t-2$) and future ($t+1$ and $t+2$) frames. Each block in frame $t$ is predicted from at least one reference frame with lower \gls{qp} (\textit{i.e.} higher quality texture information).}
    \label{fig:inter}
\end{figure}

Inter coding is mainly based on taking advantage of temporal redundancy, existing in consecutive video frames. The prediction signal in the inter mode is a block, similar to the current one, selected from within the range of \glspl{mv} search, based on a distortion metric. In modern video codecs, we are allowed to search for such similar blocks in multiple reference frames. A motion compensated signal of a given frame, defined as a composition of the most similar blocks to the blocks of the current frame, is used as the prediction information signal in the proposed method. Fig. \ref{fig:inter} visualizes how the prediction information signal is concatenated from reference frames. %which are encoded in lower \gls{qp} than the current frame. 
In this figure, current frame at time $t$ uses four reference pictures, two from the past ($t-1$, $t-2$) and two from the future ($t+1$, $t+1$).

The temporal prediction signal usually provides additional texture information which is displaced with respect to the texture in the current frame. Hence, there is a potential benefit in using the additional texture information in the temporal prediction for \gls{cnn}-based quality enhancement~\cite{nasiripcs}. Moreover, in the hierarchical \gls{gop} structure of the \acrfull{ra} and \acrfull{ld} coding configurations, these references are usually encoded using lower \gls{qp} values than that of the current frame (Fig. \ref{fig:inter}). Therefore, for some local textures of the current frame, there could occasionally be a version of the texture in a higher quality due to the lower \gls{qp} of its corresponding frame. The assumption of using the temporal prediction-aware \gls{qe} method is that feeding the prediction information to the network makes it easier to model the heavily quantized residual signal and retrieve the missing parts in the current frame. 
%The assumption of the temporal prediction-aware QE method is that feeding such higher quality textures information to the network allows it to retrieve parts of texture information that is lost in current frame, due to its relatively higher QP value 

%\hl{here what you are saying is dangerous, the QP is applied only on residual not on the predicted part from reference frame, how we can loss this quality by the QP, we can discuss if needed}. -- \hl{\textit{Is it better now}?}

\subsection{Coding Type Granularity}
\label{sec:granularity}
\subsubsection{Frame-level}
In the two previous sections, we explained how the use of prediction information could improve the performance of the \gls{qe} task. The two trained networks for intra and inter are applied differently at the frame-level. In intra frames, since all blocks have the same coding mode and prediction type, the whole frame is enhanced using the intra trained network. 

\begin{figure}
    \centering
    \begin{tabular}{cc}
        Original inter frame 
        &
        Block-type mask
        \\
        & 
        \\ 
         \includegraphics[scale=.28]{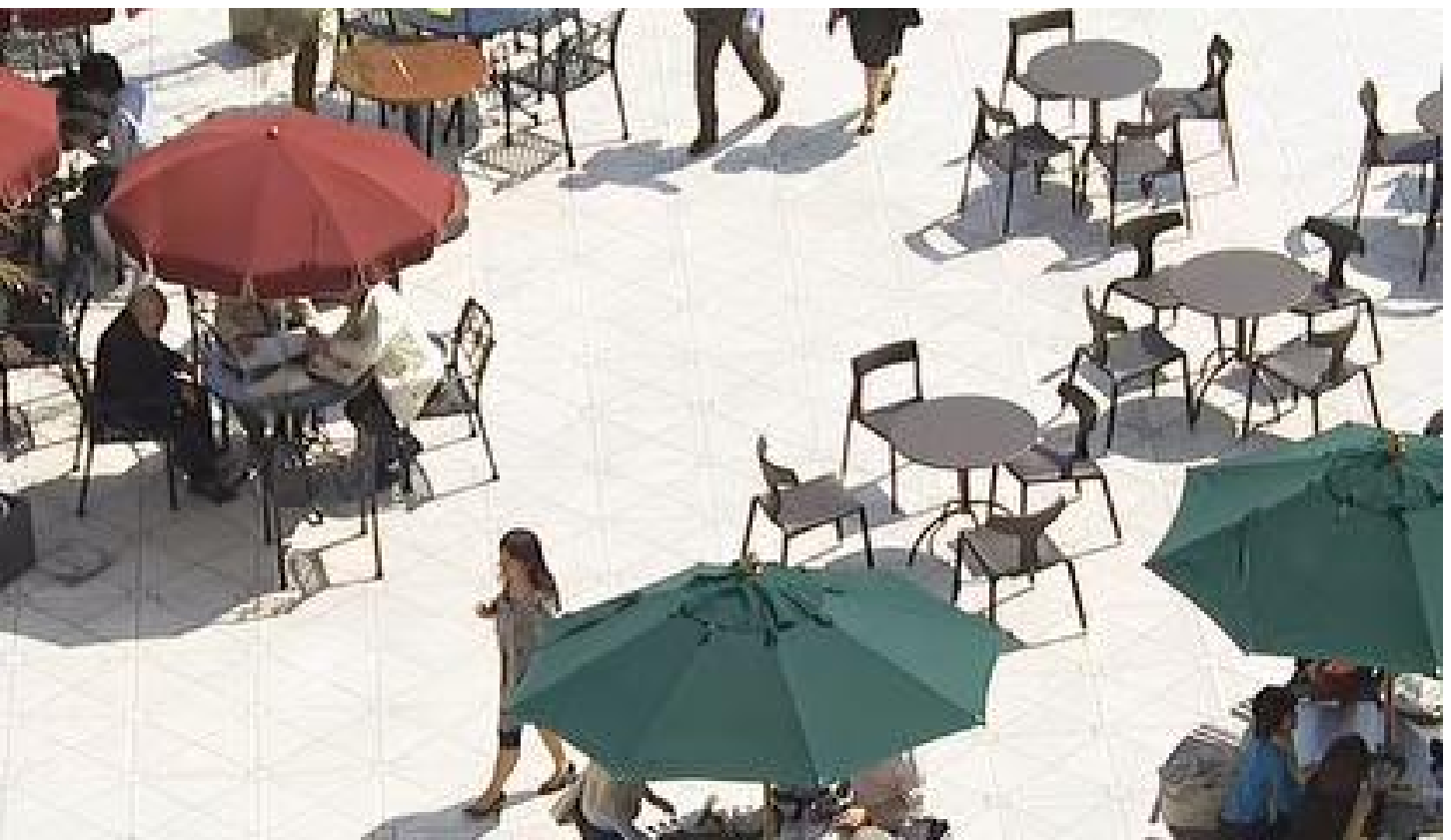}
         & 
         \includegraphics[scale=.28]{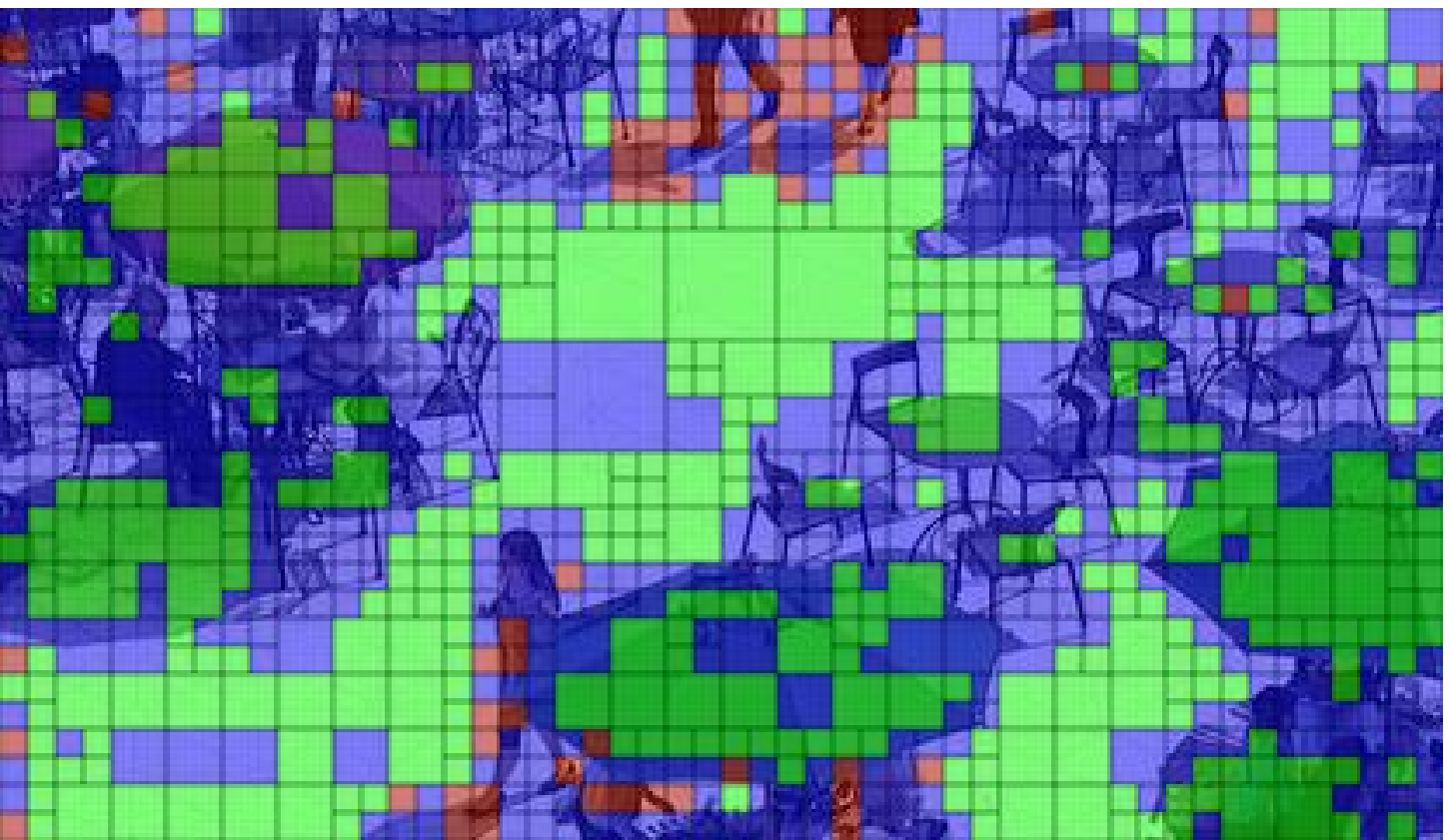}
         \\
         & 
         \\
         \multicolumn{2}{c}{\tikzset{every picture/.style={line width=0.75pt}} %set default line width to 0.75pt        

\begin{tikzpicture}[x=0.75pt,y=0.75pt,yscale=-1,xscale=1,scale=0.5]
%uncomment if require: \path (0,515); %set diagram left start at 0, and has height of 515

%Shape: Rectangle [id:dp10225531515120945] 
\draw  [fill={rgb, 255:red, 0; green, 0; blue, 255 }  ,fill opacity=1 ] (160,113) -- (198.52,113) -- (198.52,153) -- (160,153) -- cycle ;
%Shape: Rectangle [id:dp6511075700253056] 
\draw  [fill={rgb, 255:red, 170; green, 0; blue, 0 }  ,fill opacity=1 ] (100,113) -- (138.52,113) -- (138.52,153) -- (100,153) -- cycle ;
%Shape: Rectangle [id:dp3219057547039661] 
\draw  [fill={rgb, 255:red, 0; green, 255; blue, 0 }  ,fill opacity=1 ] (220,113) -- (258.52,113) -- (258.52,153) -- (220,153) -- cycle ;

% Text Node
\draw (101.75,169.29) node [anchor=north west][inner sep=0.75pt]  [font=\footnotesize] [align=left] {Intra};
% Text Node
\draw (162.75,169.29) node [anchor=north west][inner sep=0.75pt]  [font=\footnotesize] [align=left] {Inter};
% Text Node
\draw (220.75,169.29) node [anchor=north west][inner sep=0.75pt]  [font=\footnotesize] [align=left] {Skip};

\end{tikzpicture}} 
         
    \end{tabular}
    \caption{Block type mask of an inter frame from the BQSquare sequence, with the three block types present.}
    \label{fig:block_type_FR}
\end{figure}

However, in inter frames, not all blocks have the same type of prediction. 
%Depending on local texture and motion characteristics, the encoder has the choice between different types of predictions.
More precisely, three main prediction types can be found in blocks of an inter coded frame: inter, intra and skip. Fig. \ref{fig:block_type_FR} shows an example of different block types within an inter coded frame. As a result, the prediction-aware quality enhancement of inter frames is performed in the block-level.

\subsubsection{Block-level}
\label{sec:blocklevel}
The choice of the coding type of blocks in an inter coded frame depends on motion and texture characteristics. The regular inter mode, where the motion information along with residual signal is transmitted, is usually the more common type in inter coded frames. However, when the local content becomes too simple or too complex to compress, the skip mode and the intra mode might be used instead, respectively. More precisely, when a part of the video is static or has homogeneous linear motion, reference frames usually have very similar co-located blocks. In this case, skip mode is useful, where the motion is derived from neighbouring blocks and residual transmission is skipped. On contrary, due to fast motion or occlusion, sometimes no similar block can be found in reference frames. In this case, intra coding can offer a better prediction signal based on the spatial correlation of texture.

In the proposed \gls{qe} scheme, blocks within inter coded frames are enhanced based on their coding type. To do so, a block-type mask is formed using the type information extracted from the bitstream. This mask is then used to determine the proper \gls{qe} model for each block. Precisely, intra blocks and inter blocks are enhanced with intra-trained and inter-trained models, respectively. On contrary, skip blocks are enhanced using the prediction-unaware model, which is trained without any prediction information (See Eq. \ref{eq:predunaware}).
When skip mode is signalled for a block, the content of prediction signal for that block is identical to the reconstructed block. As a result, the network which is trained with inter prediction signal and has learnt the motion in the video, is unsuitable for its enhancement. Our experiments showed that if identical prediction and reconstruction signals of skip blocks are fed to the proposed prediction-aware \gls{qe} network, the performance will degrade, compared to the enhancement with the prediction-unaware method.

The implementation of the block-type mask can be performed at two levels: frame-level and block-level. In the frame-level application of the block-type mask, each inter frame is enhanced three times, using the trained networks for intra, inter and skip (prediction-unaware) coding types. Then, using the block-type mask of the frame, the three outputs are combined and one enhanced frame is produced. However, in the block-level application of the mask, the \gls{cnn}-based \gls{qe} is applied in the block-level, where each block is enhanced only once by using its appropriate model. Then all enhanced blocks are concatenated to form the final enhanced frame. Our experiments show that these two implementations have a negligible difference in terms of performance. Therefore, we chose to use the block-level approach, since it is significantly less complex in terms of the number of operations than the frame-level implementation.

%-------------------------------------------------------------------------------------------------
%-------------------------------------------------------------------------------------------------
%-------------------------------------------------------------------------------------------------

\section{Codec integration}
\label{sec:codec}
The proposed \gls{qe} method is integrated in the \gls{vvc} codec with two different approaches: \acrfull{pp} and \acrfull{ilf}. While their core \gls{qe} modules share the same principles, described in the previous section, they possess unique characteristics and impose different challenges.

\begin{figure*}
    \centering
    \input{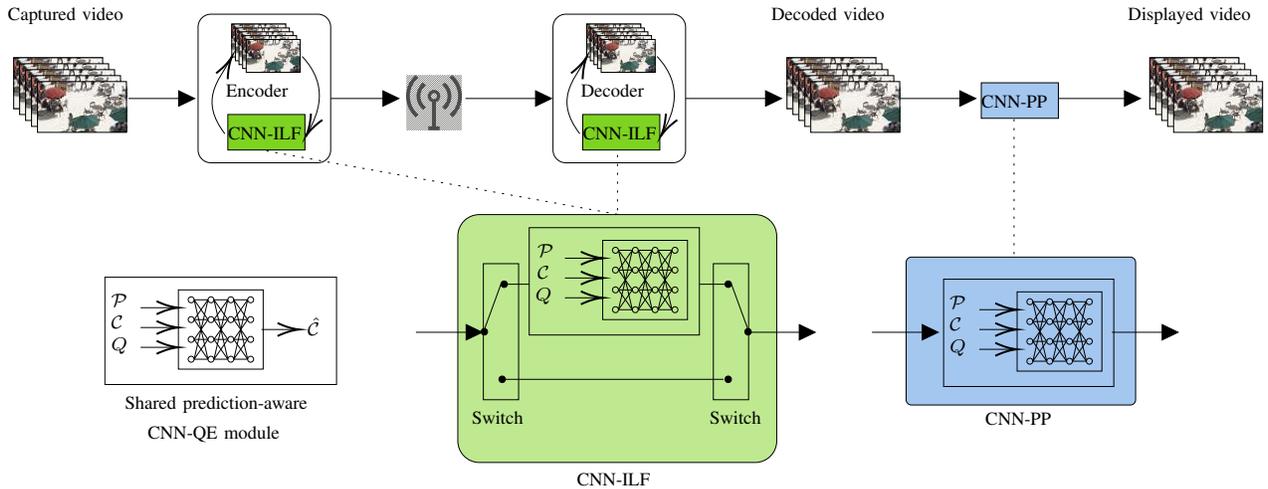}
    \caption{The proposed prediction-aware framework with two codec integration approaches: \gls{ilf} (green) and \gls{pp} (blue), sharing the same \gls{cnn}-based \gls{qe} module. }
    \label{fig:mainPP}
\end{figure*}

Fig. \ref{fig:mainPP} shows where each codec integration method is placed and how it impacts the end-to-end system. In this figure, the green and blue modules represent the \gls{ilf} and \gls{pp} approaches, respectively, where only one of them can be activated in an end-to-end system. Also, removing both of them results in the reference system where no CNN-based \gls{qe} is integrated. As can be seen, a common prediction-aware network is shared between the \gls{ilf} and \gls{pp} approaches. However, it is used differently, which will be explained later in this section.

\subsection{QE as \acrfull{pp}}
\gls{qe} as \gls{pp} module is placed after decoding the bitstream and before displaying the reconstructed image. Therefore, it is applicable only on the decoder side. From another point of view, pixel modifications of an image impact only the quality of that image with no temporal propagation. 

In this approach, the encoder side is not aware of the fact that the displayed image will go through a \gls{qe} step. Therefore, no complexity is added to the encoder and the normative aspect of the generated bitstream remains unchanged. 

At the decoder side, the \gls{pp} step is considered as optional. Usually, this choice depends on the processing capacity of the display device. For instance, if the device is equipped with dedicated \acrfull{gpu} or other neural network inference hardware, then the post-processing can be applied and bring quality improvement at no bitrate cost. Another advantage of the \gls{pp} approach is that it can be applied on already encoded videos without needing for their re-compression. 
%\hl{you can add the advantage of this approach is that it can be applied on already encoded video} -- \hl{\textit{Added. Is it OK?}}

Activating the blue box in Fig. \ref{fig:mainPP} represents the scheme of the  \gls{pp} codec integration. As the  \gls{qe} module requires the necessary coding information, namely the  \gls{qp} map, the prediction signal and the coding type mask, which are extracted from bitstream during the decoding phase. 
%\section{Experimental Results}

%-------------------------------------------------------------------------------------------------
%-------------------------------------------------------------------------------------------------
%-------------------------------------------------------------------------------------------------
\subsection{QE as In-Loop Filter (ILF)}
The main idea of an \gls{ilf} is based on the propagation of improvements. More precisely, an \gls{ilf} locally improves pixels of the current frame and then temporally propagates the improvement through frames which use the current frame as their reference. 

\gls{qe} as \gls{ilf} module is placed after existing in-loop filters in \gls{vvc} (\textit{i.e.} \gls{dbf}, \gls{sao} and \gls{alf}). Since the framework is shared with the \gls{pp} approach, same coding information is required which is accessed during the encoding process of a frame. 

Unlike the \gls{pp} approach and similar to existing \gls{vvc} in-loop filters, the \gls{ilf} approach is normative. In other words, if activated, both encoder and decoder are forced to apply it on their reconstructed samples. Therefore, one main difference of \gls{ilf} compared to the \gls{pp} is the mandatory complexity at both encoder and decoder sides. 

\begin{figure}
    \centering
    \input{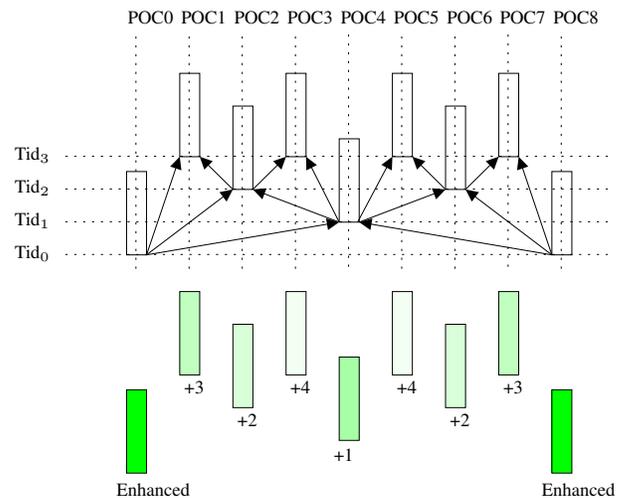}
    \caption{Propagation of quality enhancement in \gls{ilf} approach in \gls{gop} of size 8. The spectrum of greens approximately shows the benefit of each frame from the enhancement propagation, based on the distance order from the enhanced intra frames, which is approximated based on the number of steps required to reach frames from both enhanced frames. 
    %\hl{if there is any meaning of the color you should provide it in the paper} -- \hl{\textit{Added. Is it OK?}} \hl{\textit{Luce : can you say more clearly how you define the distance order?}} -- \hl{\textit{better?}}
    }
    \label{fig:ilfpropagation}
\end{figure}

On contrary, \gls{ilf} approaches have an interesting advantage of propagating the quality enhancement through the frames. Fig.~\ref{fig:ilfpropagation} visualizes this aspect. In this figure, the propagation of quality enhancement in a \gls{gop} of size 8, with four temporal layers (Tid$_i$, $i$=0,1,2,3) is shown, where only the intra frames at POC0 and POC8 are enhanced. The offsets  $\{+1, \dots ,+4\}$ approximately represents how far a frame is placed from the enhanced frames. Moreover, the spectrum of greens indicates the benefit of each frame from the enhancement propagation, based on their distance order from the enhanced intra frames. Therefore, one can see that the propagation benefit gradually diminishes as the frame gets further from the enhanced frames. 

\subsubsection{Multiple-enhancement}
The potential downside of enhancement propagation is a phenomenon called multiple enhancement in this paper. In common \gls{gop} structures with inter frames, the effect of processing one frame usually propagates through other frames that refer to it in the motion compensation. In particular, by applying in-loop quality enhancement, either \gls{cnn}-based or standard methods (e.g. \gls{sao}, \gls{alf}, \gls{dbf} etc.), when the quality of a frame in lower temporal layers is enhanced (Fig. \ref{fig:ilfpropagation}), the effective enhancement will also impact frames in higher temporal layers. For instance, a reconstructed inter frame, may contain blocks from its reference frames which are already enhanced by the applied method. 

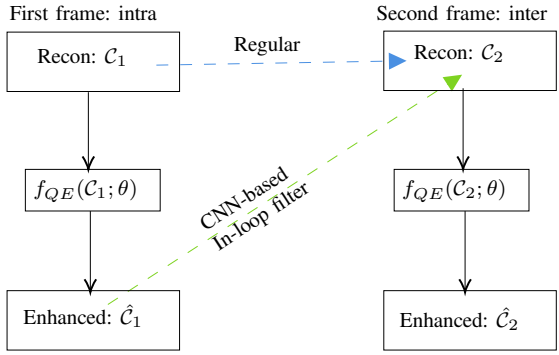
\begin{figure}
    \centering
    \tikzset{every picture/.style={line width=0.15pt}} %set default line width to 0.75pt        

\begin{tikzpicture}[x=0.75pt,y=0.75pt,yscale=-1,xscale=1]
%uncomment if require: \path (0,515); %set diagram left start at 0, and has height of 515

%Shape: Rectangle [id:dp7802037081937504] 
\draw   (110.52,91) -- (197.52,91) -- (197.52,120.98) -- (110.52,120.98) -- cycle ;
%Straight Lines [id:da9554505288283173] 
\draw    (152.52,181.33) -- (152.52,220.33) ;
\draw [shift={(152.52,222.33)}, rotate = 270] [color={rgb, 255:red, 0; green, 0; blue, 0 }  ][line width=0.75]    (6.56,-2.94) .. controls (4.17,-1.38) and (1.99,-0.4) .. (0,0) .. controls (1.99,0.4) and (4.17,1.38) .. (6.56,2.94)   ;
%Straight Lines [id:da6194145447364404] 
\draw [color={rgb, 255:red, 126; green, 211; blue, 33 }  ,draw opacity=1 ] [dash pattern={on 4.5pt off 4.5pt}]  (337.5,114.13) -- (157.52,230.98) ;
\draw [shift={(340.02,112.49)}, rotate = 147.01] [fill={rgb, 255:red, 126; green, 211; blue, 33 }  ,fill opacity=1 ][line width=0.08]  [draw opacity=0] (8.93,-4.29) -- (0,0) -- (8.93,4.29) -- cycle    ;
%Straight Lines [id:da22591585944979398] 
\draw    (340.52,119.33) -- (340.52,158.33) ;
\draw [shift={(340.52,160.33)}, rotate = 270] [color={rgb, 255:red, 0; green, 0; blue, 0 }  ][line width=0.75]    (6.56,-2.94) .. controls (4.17,-1.38) and (1.99,-0.4) .. (0,0) .. controls (1.99,0.4) and (4.17,1.38) .. (6.56,2.94)   ;
%Straight Lines [id:da07699992967935876] 
\draw    (341.52,181.33) -- (341.52,220.33) ;
\draw [shift={(341.52,222.33)}, rotate = 270] [color={rgb, 255:red, 0; green, 0; blue, 0 }  ][line width=0.75]    (6.56,-2.94) .. controls (4.17,-1.38) and (1.99,-0.4) .. (0,0) .. controls (1.99,0.4) and (4.17,1.38) .. (6.56,2.94)   ;
%Straight Lines [id:da18087068441003484] 
\draw [color={rgb, 255:red, 74; green, 144; blue, 226 }  ,draw opacity=1 ] [dash pattern={on 4.5pt off 4.5pt}]  (185.83,107.67) -- (309.02,105.54) ;
\draw [shift={(312.02,105.49)}, rotate = 539.01] [fill={rgb, 255:red, 74; green, 144; blue, 226 }  ,fill opacity=1 ][line width=0.08]  [draw opacity=0] (8.93,-4.29) -- (0,0) -- (8.93,4.29) -- cycle    ;
%Shape: Rectangle [id:dp9433582519314977] 
\draw   (300.52,90.5) -- (387.52,90.5) -- (387.52,120.48) -- (300.52,120.48) -- cycle ;
%Shape: Rectangle [id:dp21837230595645962] 
\draw   (300.52,221.5) -- (387.52,221.5) -- (387.52,251.48) -- (300.52,251.48) -- cycle ;
%Shape: Rectangle [id:dp48826496828682575] 
\draw   (110.52,221.5) -- (197.52,221.5) -- (197.52,251.48) -- (110.52,251.48) -- cycle ;
%Straight Lines [id:da04609638296346452] 
\draw    (151.52,121.33) -- (151.52,158.33) ;
\draw [shift={(151.52,160.33)}, rotate = 270] [color={rgb, 255:red, 0; green, 0; blue, 0 }  ][line width=0.75]    (6.56,-2.94) .. controls (4.17,-1.38) and (1.99,-0.4) .. (0,0) .. controls (1.99,0.4) and (4.17,1.38) .. (6.56,2.94)   ;

% Text Node
\draw (110.75,76.29) node [anchor=north west][inner sep=0.75pt]  [font=\footnotesize] [align=left] {First frame: intra};
% Text Node
\draw    (119.75,160.29) -- (187.75,160.29) -- (187.75,181.29) -- (119.75,181.29) -- cycle  ;
\draw (122.75,164.29) node [anchor=north west][inner sep=0.75pt]  [font=\footnotesize] [align=left] {$\displaystyle f_{QE}(\mathcal{C}_{1} ;\theta )$};
% Text Node
\draw (295.75,76.29) node [anchor=north west][inner sep=0.75pt]  [font=\footnotesize] [align=left] {Second frame: inter};
% Text Node
\draw (124.75,98.29) node [anchor=north west][inner sep=0.75pt]  [font=\footnotesize] [align=left] {Recon: $\displaystyle \mathcal{C}_{1}$};
% Text Node
\draw (117.75,227.29) node [anchor=north west][inner sep=0.75pt]  [font=\footnotesize] [align=left] {Enhanced: $\displaystyle \hat{\mathcal{C}}_{1}$};
% Text Node
\draw    (305.75,160.29) -- (373.75,160.29) -- (373.75,181.29) -- (305.75,181.29) -- cycle  ;
\draw (308.75,164.29) node [anchor=north west][inner sep=0.75pt]  [font=\footnotesize] [align=left] {$\displaystyle f_{QE}(\mathcal{C}_{2} ;\theta )$};
% Text Node
\draw (307.52,228.5) node [anchor=north west][inner sep=0.75pt]  [font=\footnotesize] [align=left] {Enhanced: $\displaystyle \hat{\mathcal{C}}_{2}$};
% Text Node
\draw (314.75,96.29) node [anchor=north west][inner sep=0.75pt]  [font=\footnotesize] [align=left] {Recon: $\displaystyle \mathcal{C}_{2}$};
% Text Node
\draw (224,92) node [anchor=north west][inner sep=0.75pt]  [font=\footnotesize] [align=left] {Regular};
% Text Node
\draw (205.29,184.9) node [anchor=north west][inner sep=0.75pt]  [font=\footnotesize,rotate=-327] [align=left] {CNN-based \\In-loop filter};

\end{tikzpicture}
    \caption{Multiple enhancement example in a simplified \gls{gop} of size 2 (one intra frame at left and one inter frame at right). The dashed lines show the use of reference picture for the inter frame, with (green) and without (blue) a CNN-based in-loop filter.}
    \label{fig:multienahcement}
\end{figure}

To better understand the multiple-enhancement effect, imagine a simplified low-delay \gls{gop} structure of length 2, with one intra frame and one inter frame (P-frame), as shown in Fig. \ref{fig:multienahcement}. The blue and green dashed lines from the inter frame to the intra frame represent the reference frame used for its motion compensation, without and with a \gls{cnn}-based \gls{ilf} module, respectively. The \gls{ilf} module is represented with a simplified version of Eq. \eqref{eq:summary}, where the prediction signal and the \gls{qp} map are not used in the enhancement, hence, $\mathcal{I}=\mathcal{C}$. Moreover, the reconstructed signal $\mathcal{C}$ is composed as:
\begin{equation}
    \mathcal{C=P+\hat{R}},
\end{equation}
where $\mathcal{P}$ and $\mathcal{\hat{R}}$ are the prediction signal and reconstructed residual signal (i.e. after quantization and de-quantization), respectively. Accordingly, the enhanced inter frames in Fig. \ref{eq:summary} can be expressed as:

\begin{equation}
\label{eq:multenh}
    \mathcal{\hat{C}}_2=f_{QE}(\mathcal{C}_2;\theta)=f_{QE}(\mathcal{P}_2+\mathcal{\hat{R}}_2;\theta).
\end{equation}

As often happens in content with no or limited linear motion, the residual transmission of an inter block could be skipped (i.e. the skip mode):
\begin{equation}
    \mathcal{\hat{R}}_2=0.
\end{equation}

In such circumstances, the reconstructed inter frame associated to a skip block can locally be expressed as:
\begin{equation}
   \mathcal{C}_2 = \mathcal{P}_2. 
\end{equation}

Since the prediction signal of the skip block is motion compensated from its enhanced reference, we also have: 
\begin{equation}
   \mathcal{P}_2 = \mathcal{\hat{C}}_1,
\end{equation}
which according to Eq. \eqref{eq:multenh}, it would result in multiple enhancement of the inter frame: 
\begin{equation}
    \mathcal{\hat{C}}_2 = f_{QE}(\mathcal{\hat{C}}_1;\theta)=f_{QE}(f_{QE}(\mathcal{C}_1;\theta);\theta)
\end{equation}

The multiple-enhancement effect is not an issue by nature. 
%For instance, standard \gls{ilf} methods have already taken the effect into account in their hand-crafted algorithm tuning, in order to avoid potentially harmful multiple-enhancement %\hl{how ?}.
For instance, standard \gls{ilf} also deal with a similar situation, where the reference frame has gone through the same enhancement process, and this multiple enhancement effect seems not to impact their performance.

However, in \gls{cnn}-based \gls{qe} methods, the tuning process is automatic through a complex offline training process. This aspect makes the multiple-enhancement effect a potential hazard for \gls{cnn}-based \gls{ilf} algorithms. More precisely, one of the main challenges is that a \gls{cnn}-based \gls{qe} network which is trained for the \gls{pp} task, would not perform well for the \gls{ilf} task, since it has not observed enhanced references during the \gls{pp} training. In other words, such network has observed frames whose references were not enhanced by any \gls{cnn}-based \gls{qe}. While, during the \gls{ilf} inference, this network will have to deal with reconstructed frames whose reference have also been enhanced by a \gls{cnn}-based \gls{qe}. As shown in previous studies, the multiple enhancement effect negatively impacts \gls{cnn}-based \gls{ilf} methods.

 % This may  at  first  appear  surprising  but  it  should be  rememberedthat, unlike conventional post processing, CNN-based PP does employ end-to-end training. In addition, when CNN-processed frames  are  employed  as  a  reference  (after  in-loop  filtering),they are used to predict subsequently encoded frames through motion  estimation  and  compensation.  This  process  has  not been  reflected  in  the  current  CNN  training  (i.e.  with  CNN-processed  content  as  network  input),  and  is  likely  to  cause the CNN-based filter to become less effective. Similar results have been observed by other authors when the same CNN isemployed for both PP and ILF

%\begin{figure}
%    \centering
%    \includegraphics[scale=0.13]{figs/graph_multiEnhancement.eps}
%    \caption{Caption}
%    \label{fig:multipleEnh}
%\end{figure}

\subsubsection{End-to-end training solution}
One solution to the multiple-enhancement issue is to avoid the mismatch between the training set and test set. This solution, called the end-to-end training in the literature \cite{ma2020mfrnet}, guarantees that frames whose references have been through \gls{cnn}-based \gls{qe} are present in the training set. However, since it is important that the references of such frames are also enhanced by the same \gls{cnn}-based \gls{qe}, the end-to-end training solution has a potential chicken-and-egg problem. 

One way to overcome the above problem is to run the dataset generation step and the training step in multiple iterations. Starting from the first temporal layer, in each iteration, one temporal layer is used for training and a network is trained for it. Then, all frames in that layer are enhanced in the dataset generation step, to be used in the training step of the next temporal layer. This solution is extremely time-consuming, therefore not practical for the current problem.
%Our CNN networks have been trained offline (outside of encoding process) to learn a mapping between reconstructed frame and original frame. As a result, when a frame with some blocks which are already enhanced is fed to the network as input, it will degrade the performance of the \gls{qe} network. In this case, network would process inputs which are not like the training phase inputs. In order to demonstrate the quality degradation in multiple enhancement process, we performed some experiment by applying QE filter on some specific frames. In Fig.\ref{fig:multipleEnh}, the results of in-loop enhancement while it is applied on different frames are illustrated. (I will explain it later)

\subsubsection{Adaptive ILF method}
\label{sec:switchable}
A greedy approach to avoid the multiple-enhancement is to determine at the encoder side whether or not a frame should be enhanced. For this purpose, an adaptive \gls{ilf} mechanism for inter frames is used in the proposed method. The main idea is to enhance only frames where applying the \gls{cnn}-based \gls{qe} filter results in increasing the quality. More precisely, each reconstructed frame is processed by the proposed \gls{cnn}-based \gls{qe} method at the encoder side. Then, using the original frame as reference, an \gls{mse} comparison is performed between the unprocessed reconstructed frame (before enhancement) and the processed reconstructed frame (after enhancement). The switch in the adaptive \gls{ilf} is then set based on the smaller \gls{mse} value. 

The adaptive \gls{ilf} solution requires an encoder-side signalling. Since in the proposed method, the encoder decides about the switch flag at the frame-level, the signalling is performed in the Picture Parameter Set (PPS). Signaling in the frame-level adds only one bit per frame, therefore, its impact on the coding efficiency is negligible. However, alternative implementations might apply the switch in finer granularity, such as \gls{ctu}-level or even \gls{cu}-level. This latter aspect is left as future work.

%-------------------------------------------------------------------------------------------------
%-------------------------------------------------------------------------------------------------
%-------------------------------------------------------------------------------------------------
\section{Experimental Results}
\label{sec:results}
\subsection{Experimental setup}
\subsubsection{Dataset}
The training phase has been carried out under the recent Deep Neural Network Video Coding (DNNVC) \glspl{ctc}, released by JVET \cite{JVETT2006, JVETT0129}. The recommended dataset in these \glspl{ctc} is BVI-DVC \cite{ma2020bvi}, which consists of 800 videos of 10-bit pixel representation, in different resolutions covering formats from CIF to 4K. We also used two image databases, namely DIV2K and Flickr2K for the training of intra-based networks. These datasets are composed of 900 and 2650 high quality images, respectively. The videos and images in the training dataset were converted to 10-bit YCbCr 4:2:0 and only the luma component has been used for training.

To train the models for inter frames, the native \gls{ra} configuration of VTM10.0 reference software was used with input and internal pixel depth of 10-bit. Moreover, all in-loop filters were kept activated. The video dataset was encoded in five base \glspl{qp}, \{22, 27, 32, 37, 42\}. For each \gls{qp}, four out of 64 frames of each reconstructed video were randomly selected. %In the selection process, the frames which had less number of intra coded blocks were prioritised over the other frames.
Finally, a total of 3200 reconstructed frames obtained for each \gls{qp} base, resulting in 16000 frames for all five \glspl{qp}. The equivalent ground truth and prediction signal, as well as \gls{qp} map for these reconstructed frames, were also extracted.

The networks for intra frames were trained separately by encoding the DIV2K and Flickr2K datasets in the \gls{ai} configuration. In total, 3550 images were generated, from which we randomly selected 1200 for each \gls{qp}, resulting in 5800 images. Moreover, we added the intra frames of the \gls{ra} dataset to the \gls{ai} dataset. To sum up, a total of 7400 intra frames were used. Finally, a patch-based strategy was employed which will be explained in Section \ref{TS}.  

For the test phase, nineteen sequences from the JVET \glspl{ctc} (classes A1, A2, B, C, D and E) were used \cite{JVETN1010}. It is important to note that none of these sequences were included in the training dataset. The test sequences were finally encoded in \gls{ra} and \gls{ai} configurations, using the same encoder settings as for the training. 

% All the source codes and trained models are available in ?? which allows to all reader to regenerate the results. 

%-------------------------------------------------------------------------------------------------
%-------------------------------------------------------------------------------------------------
%-------------------------------------------------------------------------------------------------

\subsubsection{Training Settings}
\label{TS}
The networks were implemented in PyTorch platform and the training was performed on NVIDIA GeForce GTX 1080Ti GPU. The parameter $N$ (number of residual blocks of the network) was set to 16. All networks were trained offline before encoding. The initial learning rate was set to $10^{-5}$ with a decay of $0.5$ for every 100 epochs. The Adam optimizer \cite{kingma2014adam} was used for back propagation during the training and each network was trained for 500 epochs. The validation dataset was extracted from the training dataset and was composed of 50 cropped reconstructed frames and their corresponding prediction and original frames. During the training, the best network parameters were chosen based on the evaluation performed on the evaluation dataset. 

 %\hl{this latter should be specified in dataset section}. 
 The training has been performed on $64 \times 64$ patches, randomly chosen from the training dataset. These patches are fed to the network on batches with a size of 16. Block rotation and flip were also applied randomly to selected patches to achieve data augmentation. 
 %For each pair of reconstructed and original patch, the corresponding prediction information is extracted. Moreover, a patch filled with the value of normalized \gls{qp} was also created. 
 
 It is important to note that one single model is shared between the three colour components ($Y$, $U$ and $V$) in all \gls{qp} values. Given that the proposed method requires different models to apply the block-level coding type mask (see Section \ref{sec:granularity}), the following models should be stored at the encoder and decoder sides:
 
 \begin{itemize}
     \item Intra-trained model, for enhancing intra frames as well as intra blocks in inter frames.
     \item Inter-trained model, for enhancing inter blocks in inter frames.
     \item Prediction-unaware model, for enhancing skip blocks in inter frames.
 \end{itemize}
 
% \hl{I think here it is confusing. It is better to say we evaluate our method in three different configurations, because only one will be integrated in real product not the three, } \\--\\ \hl{\textit{I think what I wrote was not clear. In fact, we don't have three configurations. We have only one configuration, and that is the prediction-aware configuration, which needs three models. Maybe we should discuss it tomorrow.}}. 

\subsubsection{Evaluation metrics}
The main performance metric used for comparison is the \gls{bd-br} \cite{bjontegaard2008}. This metric is formally interpreted as the amount of bit-rate saving in the same level of PSNR and VMAF based quality~\cite{vmaf}. Based on this metric, the performance of the \gls{vvc} reference software VTM-10.0 integrated with different configurations of the proposed CNN-based \gls{qe} method is presented. For this purpose, the VTM-10.0 with no modification is used as the anchor. The \gls{bd-br} saving is calculated in two ranges of \glspl{qp}, \gls{qp} $\in$ \{22, 27, 32, 37\} and \gls{qp} $\in$ \{27, 32, 37, 42\} naming \glspl{ctc} \gls{qp} range and high \gls{qp} range, respectively. 

The outputs of the tested methods are also compared in terms of \gls{psnr}. To do so, this metric is computed for each tested \gls{cnn}-based method and is noted as PSNR$_{Prop}$. Likewise, the metric is computed for the output of the reference anchor VTM-10, noted as \gls{psnr}$_{VTM}$. The original input signal before compression is used for computation of both \gls{psnr} values. The average difference between the two \gls{psnr} values, on a given set of sequences $S$ and a set of \gls{qp} values $Q$, is measured as the $\Delta$\gls{psnr} and is computed as: 
%\begin{equation}
%    \Delta \text{PSNR} = 
%    \frac{1}{l(S).l(Q)} \sum_{s \in S}\sum_{q \in Q} 
%     \left(\text{PSNR}_{Ref}^{s,q} -\text{PSNR}_{Test}^{s,q} \right),
%\end{equation}

\begin{equation}
    \Delta \text{PSNR} = \frac{\sum_{s \in S}\sum_{q \in Q} 
     \left(\text{PSNR}_{Prop}^{s,q} -\text{PSNR}_{VTM}^{s,q} \right)}{|S| |Q|}
    ,
\end{equation}
where $|S|$ and $|Q|$ denote the number of sequences and \gls{qp} values tested, respectively. Positive values of above equation indicate compression gain. In our experiments, the presented $\Delta$PSNR values are averaged over four \gls{qp} values of \glspl{ctc}.

Finally, the relative complexity of tested methods is computed as the ratio of the \gls{rt} with respect to the reference. This metric, which is applicable to both encoder and decoder sides, is computed as: 
\begin{equation}
\label{eq:complexity}
    RT = \frac{1}{|S||Q|} \sum_{s\in S} \sum_{q\in Q}  \frac{\text{RT}_{Prop}^{s,q}}{\text{RT}_{VTM}^{s,q}}
    % F = G \left( \frac{m_1 m_2}{r^2} \right)
\end{equation}

%\begin{equation}
%    Complexity = \frac{\sum_{s\in S} \sum_{q\in Q}  %\frac{\text{RT}_{Test}^{s,q}}{\text{RT}_{Ref}^{s,q}}}{len(Q).len(S)} 
%    % F = G \left( \frac{m_1 m_2}{r^2} \right)
%\end{equation}

%-------------------------------------------------------------------------------------------------
%-------------------------------------------------------------------------------------------------
%-------------------------------------------------------------------------------------------------
\begin{figure*}[!t]
    \centering
    \begin{tikzpicture}
\begin{axis}[
            ybar,
            ylabel={$\Delta$PSNR},
            symbolic x coords={BQMall, BQSquare, BQTerrace, BasketballDrill, BasketballDrive, BasketballPass, BlowingBubbles, Cactus, Campfire, CatRobot1, DaylightRoad2, FoodMarket4, FourPeople, KristenAndSara, MarketPlace, ParkRunning3, PartyScene, RaceHorses, RaceHorsesC, RitualDance, Tango2},
            xtick=data,
            bar width=2pt,
            height=4cm, 
            width=17cm, 
            enlarge x limits=0.05, 
            enlarge y limits=0.05, 
            legend style={at={(0.5,.9)},anchor=north,legend columns=-1,nodes={scale=0.8, transform shape}},
            xticklabel style = {rotate=35,anchor=east}
            %nodes near coords,
            %nodes near coords align={vertical},
            ]
\addplot[fill=green,draw=gray,opacity=.5] coordinates {(BQMall,	0.23015873	) (BQSquare,	0.558730159	) (BQTerrace,	0.116825397	) (BasketballDrill,	0.19984127	) (BasketballDrive,	0.123492063) (BasketballPass,	0.305555556	) (BlowingBubbles,	0.182857143) (Cactus,	0.131746032) (Campfire,	0.120645161) (CatRobot1,	0.141904762	) (DaylightRoad2,	0.103968254) (FoodMarket4,	0.15984127	) (FourPeople,	0.223174603	) (KristenAndSara,	0.161111111	) (MarketPlace,	0.084603175	) (ParkRunning3,	0.082380952	) (PartyScene,	0.167142857) (RaceHorses,	0.182741935	) (RaceHorsesC,	0.102096774 ) (RitualDance,	0.20047619	) (Tango2,	0.113174603)};
\addplot[fill=black,draw=gray,opacity=.5] coordinates {(BQMall,	0.268730159	) (BQSquare,	0.634603175	) (BQTerrace,	0.123809524	) (BasketballDrill,	0.232222222	) (BasketballDrive,	0.154761905) (BasketballPass,	0.364603175	) (BlowingBubbles,	0.217777778) (Cactus,	0.157619048) (Campfire,	0.148064516) (CatRobot1,	0.16984127	) (DaylightRoad2,	0.130634921) (FoodMarket4,	0.173015873	) (FourPeople,	0.231587302	) (KristenAndSara,	0.163650794	) (MarketPlace,	0.109047619	) (ParkRunning3,	0.102539683	) (PartyScene,	0.201111111) (RaceHorses,	0.23016129	) (RaceHorsesC,	0.133387097 ) (RitualDance,	0.251904762	) (Tango2,	0.146349206)};
\addplot[fill=blue,draw=gray,opacity=.5]  coordinates {(BQMall,	0.336031746	) (BQSquare,	0.710634921	) (BQTerrace,	0.181904762	) (BasketballDrill,	0.291587302	) (BasketballDrive,	0.215079365) (BasketballPass,	0.419365079	) (BlowingBubbles,	0.252857143) (Cactus,	0.208888889) (Campfire,	0.187903226) (CatRobot1,	0.225555556	) (DaylightRoad2,	0.176190476) (FoodMarket4,	0.265555556	) (FourPeople,	0.322698413	) (KristenAndSara,	0.228253968	) (MarketPlace,	0.141428571	) (ParkRunning3,	0.131111111	) (PartyScene,	0.243492063) (RaceHorses,	0.254032258	) (RaceHorsesC,	0.162580645 ) (RitualDance,	0.346825397	) (Tango2,	0.198412698)};
\addplot[fill=red,draw=gray,opacity=.5]   coordinates {(BQMall,	0.331111111	) (BQSquare,	0.65015873	) (BQTerrace,	0.19015873	) (BasketballDrill,	0.296349206	) (BasketballDrive,	0.215079365) (BasketballPass,	0.416031746	) (BlowingBubbles,	0.252539683) (Cactus,	0.224920635) (Campfire,	0.195483871) (CatRobot1,	0.241269841	) (DaylightRoad2,	0.196190476) (FoodMarket4,	0.294444444	) (FourPeople,	0.332063492	) (KristenAndSara,	0.179365079	) (MarketPlace,	0.151428571	) (ParkRunning3,	0.134603175	) (PartyScene,	0.218253968) (RaceHorses,	0.257580645	) (RaceHorsesC,	0.169677419 ) (RitualDance,	0.355396825	) (Tango2,	0.215873016)};

\legend{$N$=\text{4\hspace{10pt}},$N$=\text{8\hspace{10pt}},$N$=\text{16 (proposed)\hspace{10pt}},$N$=\text{32\hspace{10pt}}}
\pgfplotsset{every x tick label/.append style={font=\scriptsize}}
\end{axis}
\end{tikzpicture}
    \caption{Impact of the number of residual layers ($N$) on the performance of the network in terms of $\Delta$PSNR. The reference for this test is VTM10, hence positive $\Delta$PSNR values indicate higher quality enhancement. All test are carried out in the \gls{ra} mode.
    %\hl{what is the refrence what is the QP + is gain of loss} -- \hl{\textit{Is it OK now?}}
    } 
    \label{fig:netcomplexity}
\end{figure*}
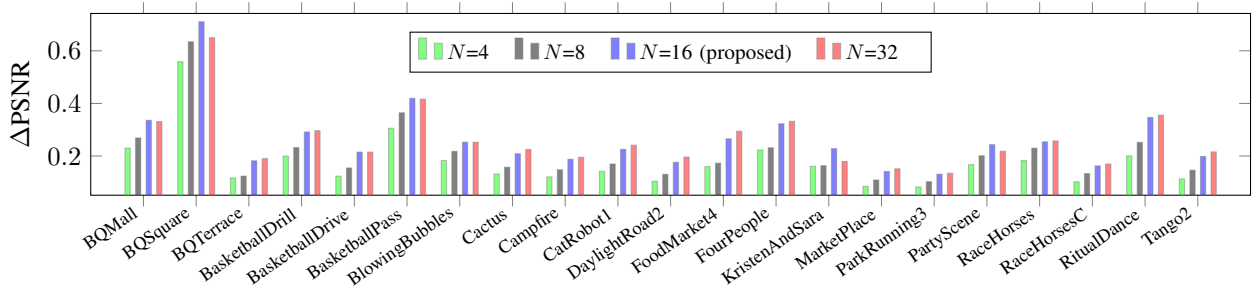

\subsection{Ablation Study}
In this section, the impact of the following elements of the proposed method are analysed: \gls{qp} map training, network depth and prediction-awareness. It should be noted that for this ablation study, results and discussions are limited only to the \acrfull{pp} integration of the proposed method. This way, different elements of the proposed method can be evaluated without entering into the complexity of multiple-enhancement and temporal aspect of the \acrfull{ilf} integration.

\subsubsection{Network Architecture}
Generally, more complex network architectures have a larger capacity of learning complicated tasks, such as coding artifacts. In the first ablation study, we modify the network architecture, presented in Fig. \ref{arch}, in terms of the number of residual blocks $N$. More precisely, instead of using $N$=16, we try alternative values 4, 8 and 32. Fig. \ref{fig:netcomplexity} shows that decreasing the parameter $N$ impacts the \gls{qe} performance (\gls{ra} configuration) in terms of $\Delta$PSNR. As can be seen, the global trend is a decrease in the performance. However, the difference between $N$=16 and $N$=32 is negligible. Therefore, in this paper we chose $N$=16 for the network architecture.
%\hl{The legend of the Figure is strange also several missing parameters which qp average ?} \\ -- \\\hl{I explained for Eq (13) that all $\Delta$PSNR values are average of 4 QPs. Is it enought or I precise here again?}. 

\subsubsection{\gls{qp}-map training}
\gls{qp} has been used as an input to all \gls{qe} networks presented in this paper. However, a noticeable number of studies in the literature take the \gls{qp}-specific training approach, assuming that several trained networks can be stored at the encoder and/or decoder side. 

A set of ablation studies have been conducted to understand differences between the two approaches in the \acrfull{ai} coding mode. To this end, in addition to performance comparison of the two approaches, the potential damage due to the use of the incorrectly trained network in the \gls{qp}-specific approach is also studied. More precisely, each network which has been trained on a particular \gls{qp} was used for the \gls{qe} task of other \gls{qp} values. 

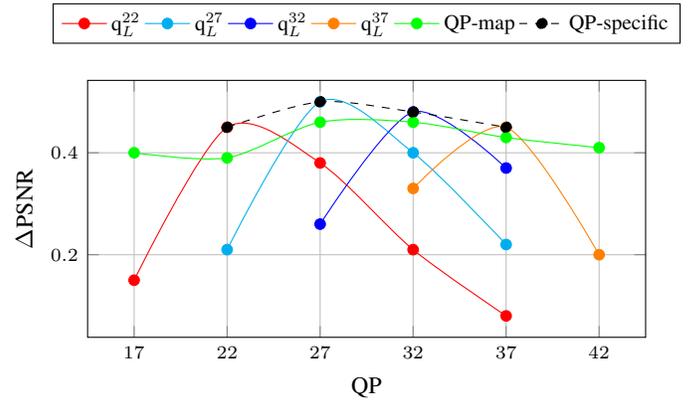
\begin{figure}
    \centering
    \begin{tikzpicture}
\begin{axis}[
	xlabel={QP},
	ylabel={$\Delta$PSNR},
	width=9cm,
	height=5cm,
	label style={font=\small},
	grid=major,
	legend style={at={(0.5,1.3)},anchor=north,legend columns=-1,nodes={scale=0.8, transform shape}},
	xtick={17,22, 27, 32, 37,42},
]

% 22
\addplot[smooth, mark=*, color=red]    coordinates 
    {(17, .15) (22, .45) (27, .38)   (32, .21) (37, .08) 
    %(42, 0) 
    };
    
% 27 
\addplot[smooth, mark=*, color=cyan]    coordinates 
    {
    %(17, 0) 
    (22, .21) (27, .5)   (32, .4) (37, .22) 
    %(42, 0) 
    };

% 32
\addplot[smooth, mark=*, color=blue]    coordinates 
    {
    %(15, 0)  (22, 0) 
    (27, .26)   (32, .48) (37, .37) 
    %(42, 0)
    };

% 37
\addplot[smooth, mark=*, color=orange] coordinates 
    {
    %(15, 0) (22, 0)    (27, 0)   
    (32, 0.33) (37, .45) (42, .2)};
    
% qp-map
\addplot[smooth, mark=*, color=green] coordinates 
    {(17, .4) (22, 0.39)    (27, 0.46)   (32, 0.46) (37, .43) (42, .41)};
    
% qp-specific
\addplot[smooth, mark=*, color=black, dashed] coordinates 
    {(22, 0.45)    (27, 0.5)   (32, 0.48) (37, .45)};
    
%\addplot[smooth, mark=*, color=green, dashed] coordinates 
%    {(17, .4) (22, 0.45)    (27, 0.46)   (32, 0.46) (37, .43) (42, .41)};

\legend{q$_L^{22}$,q$_L^{27}$,q$_L^{32}$,q$_L^{37}$,QP-map, QP-specific}
\pgfplotsset{every x tick label/.append style={font=\scriptsize}}
\pgfplotsset{every y tick label/.append style={font=\scriptsize}}
\end{axis}
\end{tikzpicture}
    \caption{PSNR performance of the \gls{qp}-specific and \gls{qp}-map training. %Moreover, the performance of \acrshort{cnn} models of the \acrshort{qp}-specific mode when they are used on different \acrshort{qp} values (\glspl{ctc} range) are presented. 
    The reference for the $\Delta$PSNR computation is VTM-10 and all test are carried out in the \gls{ai} mode.}
    \label{fig:qpmap}
\end{figure}

The result is presented in Fig. \ref{fig:qpmap}, where the average $\Delta$PSNR with respect to VTM-10 is used as metric (computed on all \glspl{ctc} sequences). In this figure, the proposed method based on the \gls{qp}-map (shown in green) is compared to five other configurations. Four of them, expressed as q$_L^i$ with $i=$22, 27, 32 and 32, are the configurations where one single model trained on the \gls{qp} value $i$ is used for enhancement of all other \gls{qp}s. As the fifth one, each of above models are used for their exact \gls{qp} value, which results in the  \gls{qp}-specific methods in the literature (shown with a dashed black curve). 

As can be seen, the use of \gls{qp}-specific training approach is slightly better than the \gls{qp}-map approach. However, the cost of storing several models makes it less useful from the implementation point of view. Especially, this additional cost is problematic at the decoder side which is supposed to be implemented various devices with different range of capacities, including mobile devices with considerably limited hardware resources. Moreover, it can be seen that the \acrshort{qp}-map training approach has a robust performance when applied on different \gls{qp} values. On the contrary, the models that are trained for a particular \gls{qp} usually have a poor performance on any other \gls{qp} value, therefore, sharing them for a range of \gls{qp} values can also damage the performance.

\subsubsection{Prediction-awareness}
As the main contribution, the prediction-awareness of the proposed method is evaluated in Table \ref{tab:results}, in terms of \gls{bd-br} gain compared to VTM-10. 
%\hl{you should say what the table provide B D-BR  CTC ?} -- \hl{\textit{Fixed?}} 
For this table, the proposed method is integrated as the \gls{pp} module in the \gls{ra} coding configuration. This means that both intra and inter frames have been enhanced using their dedicated networks. Moreover, the \gls{bd-br} metric is measured in two ranges of \glspl{ctc} and high \glspl{qp}. Finally, two configurations of the proposed \gls{qe} framework are evaluated on luma and chroma components: prediction-unaware (described in Eq. \eqref{eq:predunaware}) and prediction-aware (described in Eq. \eqref{eq:predaware}). The benchmark for the \gls{bd-br} comparison is the prediction-unaware version of the proposed \gls{cnn}-based \gls{qe} method. 

As can be seen, the proposed prediction-aware algorithm consistently outperforms the prediction-unaware one, in both \gls{qp} ranges. In the \glspl{ctc} \gls{qp} range, the coding gains (\gls{bd-br}(PSNR)) of prediction-awareness on $Y$, $U$ and $V$ are -7.31\% -8.90\% -11.22\%, respectively. Compared to the prediction-unaware setting with coding gains of -5.79\%, -8.11\%, -9.53\%, it can be noticed that adding prediction-awareness brings -1.52\%, -0.79\% and -1.69\% more bitrate savings in the three components, respectively. Moreover, the VMAF-based \gls{bd-br} results show even higher performance improvement when prediction information is used. Precisely, the gains for prediction-aware and prediction-unaware setting are -9.2\% and -5.5\%, respectively, which indicates -3.7\% more bit-rate saving in terms of VMAF-based \gls{bd-br}.

Likewise, a consistent luma \gls{bd-br} gain of 1.03\% by the prediction-aware compared to the prediction-unaware method can be observed in the high \gls{qp} range. The relative gain in the high \gls{qp} range is about 0.5\% smaller than in the \glspl{ctc} range. This could be explained by the fact that the absolute gains of both prediction-unaware and prediction-aware methods are larger in this \gls{qp} range than in the \glspl{ctc} range, possibly causing a saturation of performance gain.

%\hl{you should first describe the number of the table versus the anchors then the differences; not clear otherwise}. 
   
\begin{table*}[!ht]
\caption{\acrshort{bd-br} metric for performance comparison of the proposed \gls{cnn}-based \gls{qe} method as Post-Processing in the RA coding configuration on top of VTM-10.0.}
\label{tab:results}
\centering
\scalebox{0.96}{% Please add the following required packages to your document preamble:
\renewcommand{\arraystretch}{1.15} 
\begin{tabular}{P{0.007\textwidth}C{0.07\textwidth}cccc|cccc||cccc|cccc}
\hline
\hline							
 \multirow{4}{*}{Class } & \multirow{4}{*}{ Sequence} & \multicolumn{8}{c||}{\glspl{ctc} QP (22-37)}	& \multicolumn{8}{c}{High QP (27-42)}\\ \cline{3-18}                       

                       &                     & \multicolumn{4}{c|}{Prediction-unaware}                      & \multicolumn{4}{c||}{Prediction-aware}               & \multicolumn{4}{c|}{Prediction-unaware}                                  & \multicolumn{4}{c}{Prediction-aware} \\
					   \cline{3-18}
					                          &                     & \multicolumn{3}{c}{BD-BR (PSNR)} & BD-BR                     & \multicolumn{3}{c}{BD-BR (PSNR)} & BD-BR               & \multicolumn{3}{c}{BD-BR (PSNR)} & BD-BR                                  & \multicolumn{3}{c}{BD-BR (PSNR)} & BD-BR \\

\cline{3-5} \cline{7-9} \cline{11-13} \cline{15-17} 
                       &                        & Y                      & U                      & V          & (VMAF)          & Y                      & U                      & V           & (VMAF)           
					   & Y                      & U                      & V                      & (VMAF) & Y                      & U                      & V               & (VMAF)       \\
\hline
\multirow{4}{*}{A1} 	&	Tango	        &	-5.6 	&	-14.4	&	-12.5	&	-6.4  & -8.5	&	-15.7	&	-14.5	& -11.7 &	-6.1	&	-12.7	&	-11.7	& -6.8 &	-7.9	&	-13.6	&	-12.8	& -11.8 \\	
                    	&	FoodMarket	    &	-3.6	&	-11.4	&	-8.4	&	-11.9 & -7.2	&	-11.5	&	-10.5	& -15.5 &	-5.8	&	-11.6	&	-10.0	& -12.0&	-7.7	&	-11.7	&	-11.0	& -15.5 \\	
                    	&	CampFire	    &	-4.3	&	-4.2	&	-10.8	&	-9.9  & -5.9	&	-6.9	&	-14.2	& -14.2 &	-5.9	&	-4.3	&	-10.6	& -9.1 &	-7.4	&	-6.8	&	-13.6	& -12.7 \\	
                    	&	Average	        &	-4.5	&	-10.0	&	-10.6	&	-9.4  & -7.2	&	-11.4	&	-13.0	& -13.8 &	-5.9	&	-9.5	&	-10.8	& -9.3 &	-7.7	&	-10.7	&	-12.4	& -13.3 \\	\hline
\multirow{4}{*}{A2} 	&	CatRobot	    &	-6.9	&	-13.3	&	-12.8	&	-5.4  & -8.5	&	-13.6	&	-13.5	& -11.1 &	-6.8	&	-11.0	&	-11.5	& -5.7 &	-7.8	&	-11.2	&	-11.2	& -10.9 \\	
                    	&	Daylight	    &	-9.3	&	-11.3	&	-6.1	&	-9.4  & -10.8	&	-11.5	&	-8.6	& -15.4 &	-8.1	&	-8.8	&	-3.9	& -9.7 &	-9.0	&	-8.8	&	-5.1	& -15.4 \\	
                    	&	ParkRunning	    &	-3.0	&	-2.6	&	-3.5	&	-1.4  & -4.1	&	-2.4	&	-4.2	& -5.5  &	-3.2	&	-2.7	&	-3.4	& -1.8 &	-4.2	&	-2.9	&	-3.9	& -5.3  \\	
                    	&	Average	        &	-6.4	&	-9.1	&	-7.5	&	-5.4  & -7.8	&	-9.2	&	-8.8	& -10.7 &	-6.1	&	-7.5	&	-6.3	& -5.7 &	-7.0	&	-7.6	&	-6.7	& -10.6 \\	\hline
\multirow{6}{*}{B}  	&	MarketPlace	    &	-4.8	&	-7.2	&	-8.5	&	-5.0  & -5.6	&	-8.0	&	-9.8	& -7.5  &	-4.6	&	-5.8	&	-9.2	& -5.2 &	-5.2	&	-6.2	&	-10.0	& -7.3  \\	
                    	&	RitualDance	    &	-6.0	&	-9.0	&	-11.3	&	-8.5  & -8.0	&	-10.6	&	-13.4	& -11.7 &	-6.2	&	-8.0	&	-12.5	& -8.3 &	-7.6	&	-9.0	&	-13.8	& -11.2 \\	
                    	&	Cactus	        &	-4.2	&	-6.3	&	-8.5	&	-6.3  & -6.0	&	-6.3	&	-10.2	& -8.6  &	-5.5	&	-5.9	&	-9.3	& -6.4 &	-6.7	&	-6.0	&	-10.3	& -8.5  \\	
                    	&	BasketballDrive	&	-5.4	&	-6.8	&	-13.4	&	-3.9  & -7.0	&	-10.1	&	-15.3	& -7.3  &	-5.8	&	-2.2	&	-13.4	& -4.7 &	-7.0	&	-9.4	&	-14.4	& -7.7  \\	
                    	&	BQTerrace	    &	-4.9	&	-11.8	&	-9.9	&	1.3   & -5.9	&	-13.3	&	-12.8	& -2.9  &	-6.7	&	-7.7	&	-6.6	& 0.1  &	-7.4	&	-8.4	&	-7.6	& -3.3  \\	
                    	&	Average	        &	-5.1	&	-8.2	&	-10.3	&	-4.5  & -6.5	&	-9.7	&	-12.3	& -7.6  &	-5.8	&	-5.9	&	-10.2	& -4.9 &	-6.8	&	-7.8	&	-11.2	& -7.6  \\	\hline
\multirow{5}{*}{C}  	&	BasketballDrill	&	-6.5	&	-12.0	&	-15.3	&	-6.4  & -8.3	&	-12.3	&	-16.3	& -9.4  &	-6.6	&	-10.2	&	-16.3	& -6.4 &	-7.9	&	-10.8	&	-15.8	& -9.2  \\	
                    	&	BQMall	        &	-5.2	&	-5.2	&	-6.4	&	-4.9  & -6.7	&	-5.5	&	-7.4	& -8.7  &	-6.4	&	-3.7	&	-6.6	& -5.0 &	-7.2	&	-4.0	&	-6.9	& -8.4  \\	
                    	&	PartyScene	    &	-5.3	&	-4.7	&	-7.3	&	-3.9  & -6.1	&	-4.8	&	-8.2	& -7.5  &	-5.9	&	-3.6	&	-6.6	& -4.0 &	-6.3	&	-3.8	&	-6.7	& -7.0  \\	
                    	&	RaceHorses	    &	-2.8	&	-8.4	&	-8.6	&	-4.0  & -4.2	&	-9.4	&	-11.6	& -7.4  &	-3.7	&	-9.0	&	-9.4	& -4.1 &	-4.9	&	-9.7	&	-11.7	& -7.1  \\	
                    	&	Average	        &	-5.0	&	-7.6	&	-9.4	&	-4.8  & -6.3	&	-8.0	&	-10.9	& -8.2  &	-5.6	&	-6.6	&	-9.7	& -4.9 &	-6.6	&	-7.1	&	-10.3	& -7.9  \\	\hline
\multirow{5}{*}{D}  	&	BasketballPass	&	-8.0	&	-7.5	&	-16.6	&	-7.2  & -8.9	&	-7.9	&	-17.2	& -10.0 &	-9.0	&	-6.3	&	-16.8	& -6.7 &	-9.4	&	-6.8	&	-16.8	& -9.2  \\	
                    	&	BQSquare	    &	-12.4	&	-3.8	&	-5.9	&	1.1   & -12.8	&	-4.3	&	-6.8	& -2.4  &	-12.8	&	-0.5	&	-3.7	& 0.1  &	-12.9	&	-1.2	&	-4.2	& -2.8  \\	
                    	&	BlowingBubble	&	-6.2	&	-5.3	&	-6.6	&	-7.3  & -7.0	&	-5.9	&	-8.4	& -9.2  &	-6.6	&	-3.6	&	-4.7	& -6.7 &	-7.5	&	-4.3	&	-6.2	& -8.4  \\	
                    	&	RaceHorses	    &	-5.6	&	-8.8	&	-8.6	&	-6.0  & -7.4	&	-9.0	&	-10.4	& -8.3  &	-5.7	&	-7.9	&	-8.8	& -5.3 &	-7.1	&	-8.3	&	-9.3	& -7.3  \\	
                    	&	Average	        &	-8.0	&	-6.4	&	-9.4	&	-4.9  & -9.0	&	-6.8	&	-10.7	& -7.5  &	-8.5	&	-4.6	&	-8.5	& -4.7 &	-9.2	&	-5.2	&	-9.1	& -6.9  \\	\hline
					\multicolumn{2}{c}{\textbf{All}}	&   \textbf{-5.8}	&	\textbf{-8.1}	&	\textbf{-9.5}	&	\textbf{-5.5}  & \textbf{-7.3}	&	\textbf{-8.9}	&	\textbf{-11.2}   & \textbf{-9.2}  &	\textbf{-6.4}	&	\textbf{-6.6}	&	\textbf{-9.2}	& \textbf{-5.7} &	\textbf{-7.4}	&	\textbf{-7.5}	&	\textbf{-10.1} & \textbf{-8.9}
\\ 
\hline
\hline
\end{tabular}
}
\end{table*}

%-------------------------------------------------------------------------------------------------
%-------------------------------------------------------------------------------------------------
%-------------------------------------------------------------------------------------------------
\begin{table*}[!ht]
    \caption{\acrshort{bd-br} comparison of the proposed method against state-of-the-art \gls{pp} methods. All tests have been carried out in the \acrshort{ra} mode and under JVET-\glspl{ctc}.}
    \centering
    \renewcommand{\arraystretch}{1.15} 
\begin{tabular}{P{0.05\textwidth}P{0.1\textwidth}P{0.1\textwidth}P{0.1\textwidth}P{0.1\textwidth}P{0.1\textwidth}|P{0.1\textwidth}P{0.1\textwidth}}
\hline
\hline

\multirow{3}{*}{Class}	& \multicolumn{5}{c|}{ State-of-the-art}					 & \multicolumn{2}{c}{Proposed} \\ 
\cline{2-8} 
						 &JVET-O0132 \cite{JVETO0132}        & JVET-O0079 \cite{JVETO0079}   & Zhang \textit{et al.} \cite{zhang2020enhancing}        & JVET-T0079  \cite{JVETT0079} & MFRNet \cite{ma2020mfrnet}   & Pred-unaware      & Pred-aware \\
						& VTM-4			& VTM-5		          & VTM-4				  				& VTM-10				   &VTM-7			                  & VTM-10		              & VTM-10			     		\\

\hline	                  
A1    				   	& -0.15\%                   & -0.87\%          & -2.41\%                            & -2.86\%                      & -6.73\%                 & -4.47\%            & -7.20\%                        \\  						
A2    					& -0.28\%                   & -1.68\%          & -4.22\%                            & -2.98\%                      & -7.16\%                  & -6.41\%            & -7.79\%                        \\  
B     					& -0.22\%                   & -1.47\%          & -2.57\%                             & -2.92\%                     & -6.30\%                  & -5.06\%            & -6.50\%                        \\  
C     					& -0.59\%                   & -3.34\%          & -3.89\%                             & -2.96\%                     & -6.00\%                  & -4.97\%            & -6.35\%                       \\  
D     					& -0.80\%                   & -4.97\%          & -5.80\%                             & -3.48\%                     & -7.60\%                  & -8.04\%            & -9.01\%                        \\  
\hline                                                                                                                                                    
\textbf{All}            & \textbf{-0.40\%}          & \textbf{-2.47\%} & \textbf{-3.76\%}                    & \textbf{-3.04\%}             & \textbf{-6.70\%}          & \textbf{-5.79\%}   & \textbf{-7.31\%}              \\ 
\hline
\hline
%\makegapedcells
\end{tabular}

    \label{tab:stateoftheartpp}
\end{table*}

\subsection{Performance evaluation of PP}
In this section, performance of a set of recent \gls{pp} methods developed for \gls{ra} of \gls{vvc} are compared to our proposed method. For this purpose, two academic papers \cite{ma2020mfrnet,jin2020post} and three JVET contributions \cite{JVETT0079, JVETO0132, JVETO0079}, have been selected. The coding gain of these works is extracted from its corresponding literature. It is important to note that a fair comparison of \gls{cnn}-based \gls{qe} methods in the literature is difficult since they use a network with different architecture and complexity levels. 

The performance of five above-mentioned methods, as well as our proposed \gls{qe}, in terms of \gls{bd-br} are summarized in Table \ref{tab:stateoftheartpp}. The average \gls{bd-br} of each class is shown for comparison. First, it can be observed that when our proposed \gls{qe} is integrated as \gls{pp} to the \gls{vvc}, it outperforms all the competing methods. The performance improvement is consistent over the average of all classes. Secondly, the coding gain of our prediction unaware setting is less than the MFRNet \cite{ma2020mfrnet}. It can be concluded that, by adding the same strategies to the network of MFRNet, we can even get higher coding gain. This subject will be studied in future. Finally, the work in JVET-T0079 \cite{JVETT0079} also benefits from intra prediction for enhancing the quality of intra coded frames in \gls{ra} configuration. The higher \gls{bd-br} in our method compared to this work is likely due to the inter prediction and coding type mask that is employed in the proposed \gls{qe} for enhancing the inter coded frames.

%\hl{The comparison is not fair. Do you know how we can write a fair analysis?} \hl{can you draw a Figure with BD-rate vs number of parameters I can give you some examples}

%-------------------------------------------------------------------------------------------------
%-------------------------------------------------------------------------------------------------
%-------------------------------------------------------------------------------------------------

\subsection{Performance evaluation of ILF}
In terms of complexity-performance trade-off, the main benefit of the \gls{ilf} approach is that, due to the propagation of improvement, one can achieve higher compression gain by enhancing a few frames which are referred to the most in the given \gls{gop} structure. In this section, we evaluate this aspect. For this reason, here a set of \acrshort{ilf} \acrshort{qe} configurations are defined for presenting the performance in different conditions. %\hl{all the previous sectiosn it was not clear if you consider the mask block levels frame level ?}. -- \hl{\textit{I changed Section III-C (Coding Type Granularity) to explain that we apply both in the frame-level and block-level. Frame-level for intra frames and block-level for inter frames. Is it better?}}

%-------------------------------------------------------------------------------------------------
%\subsubsection{\gls{gop} Configurations and comparison setting}
\subsubsection{\acrshort{ilf} \acrshort{qe} configurations}
\label{sec:ILFgopconfigs}
%\hl{you should say that in this configuration with dont use the native VVC RA conf but 8 GOP presented in Fig ...} 

The use of sixteen frames in one \gls{gop}, as recommended in the native \gls{ra} configuration of VTM, results in five temporal layers Tid$_i$, with $0 \leq i \leq 4$. For the experiments of this section, we define six \acrshort{ilf} \acrshort{qe} configurations, to progressively increase the number of enhanced frames in the \gls{ilf} approach. In the first configuration, noted as $C_{I}$, only the intra frame in the \gls{gop} is enhanced. In each of the other five configurations, represented as $C_j$, with $0\leq j \leq 4$, all frames in the temporal layers Tid$_i$ (with $i<j$) are enhanced (Table \ref{tab:gopconfig1}).

\begin{table}
\centering
\renewcommand{\arraystretch}{1.15} 
\caption{Description of the tested \acrshort{ilf} \acrshort{qe} configurations used for evaluation of the \acrshort{ilf} approach. In each configuration, frames in some temporal layer of the \gls{gop} are enhanced (\cmark) and some are not enhanced (\xmark). All tests are carried out in the \acrshort{ra} mode.}
\begin{tabular}{cccccccc}

\hline
\hline
\multicolumn{1}{c}{Temporal}    & \multicolumn{7}{c}{\acrshort{ilf} \acrshort{qe} configuration} \\
\cline{2-8}
\multicolumn{1}{c}{layer ID}      & $Ref$ & $C_I$ & $C_0$ & $C_1$ & $C_2$ & $C_3$ & $C_4$  	\\
\hline
Intra   	& \xmark 	& 	\cmark	& 	\cmark	& 	\cmark	& 	\cmark	& 	\cmark	& 	\cmark		\\ 
0       	& \xmark 	& 	\xmark	& 	\cmark	& 	\cmark	& 	\cmark	& 	\cmark	& 	\cmark		\\ 
1       	& \xmark 	& 	\xmark	& 	\xmark	& 	\cmark	& 	\cmark	& 	\cmark	& 	\cmark		\\ 
2       	& \xmark 	& 	\xmark	& 	\xmark	& 	\xmark	& 	\cmark	& 	\cmark	& 	\cmark		\\ 
3       	& \xmark 	& 	\xmark	& 	\xmark	& 	\xmark	& 	\xmark	& 	\cmark	& 	\cmark		\\ 
4       	& \xmark 	& 	\xmark	& 	\xmark	& 	\xmark	& 	\xmark	& 	\xmark	& 	\cmark		\\ 
\hline
\hline

\end{tabular}
\label{tab:gopconfig1}
\end{table}

%Earlier, a \gls{gop} of size eight has been discussed in Fig. \ref{fig:ilfpropagation}. However, the common \gls{gop} size of \gls{vvc} consists of sixteen frames, resulting in five temporal layers Tid$_i$, with $0 \leq i \leq 4$. 
%\hl{it is not realy an equation}: 

%\begin{equation}
%    \label{eq:gopconfig}
%    C_{j} : \text{All Tid}_i \text{ with } i<j \text{ are enhanced.}
%\end{equation}

%The selected \gls{gop} configurations for our experiments are $C_{0}$, $C_{1}$, $C_{2}$, $C_{3}$, $C_{4}$ and $C_{I}$. 
Additionally, we also present a configuration as $C_{ref}$ which is equivalent to the VTM-10 encoder with no CNN-based \gls{qe} and is included to be used as a reference. Using the defined \acrshort{ilf} \acrshort{qe} configurations, we present the results of the proposed \gls{ilf} method in a progressive manner so that the impact of multiple-enhancement is reflected. For this purpose, four settings of the VTM encoder are evaluated: 

\begin{itemize}
    \item Prediction-aware \gls{ilf}.
    \item Prediction-unaware \gls{ilf}.
    \item Adaptive \gls{ilf} (prediction-aware only).
    \item Prediction-aware \gls{pp}.
\end{itemize}

%-------------------------------------------------------------------------------------------------
\subsubsection{Performance Evaluation of ILF QE configurations}
\label{sec:ILFconfigs}
\begin{figure}
    \centering
    \begin{tikzpicture}
\begin{axis}[
	xlabel={ILF QE Configuration},
	%no markers,
	xticklabel pos=left,
	%axis x line*=top,
	ytick = {-6, -3, 0, +3, +6},
	ylabel={BD-BR\%},
	width=8.5cm,
	height=6cm,
	grid=major,
	legend style={at={(0.02,.655)},anchor=south west, nodes={scale=0.7, transform shape}},
	symbolic x coords={minus,$C_{ref}$,$C_{I}$,$C_{0}$,$C_{1}$,$C_{2}$,$C_{3}$,$C_{4}$,plus},
    xtick=data,
]

%ilf-fusion1-sw0	
\addplot coordinates {($C_{ref}$, 0) ($C_{I}$,	-3.757519577)	($C_{0}$,	-5.09658087	)	($C_{1}$,	-5.21019667 )	($C_{2}$,	-3.868692388)	($C_{3}$,	-0.579577885)	($C_{4}$, 7.1)	                     };

%ilf-fusion0-sw0	
\addplot coordinates {($C_{ref}$, 0) ($C_{I}$,	-3.603979826)	($C_{0}$,	-4.627674305	)	($C_{1}$,	-4.28416821 )	($C_{2}$,	-2.889588079)	($C_{3}$,	0.157032964)	($C_{4}$, 6.95359906)	                     };

%ilf-fusion1_sw1	
\addplot [smooth, color=green, dashed] coordinates {($C_{I}$,	-5.7)	($C_{0}$,	-5.7)	($C_{1}$,	-5.7)	($C_{2}$,	-5.7)	($C_{3}$,	-5.7)	($C_{4}$,	-5.7)};

%pp-fusion0		
\addplot coordinates {($C_{ref}$, 0) ($C_{I}$,	-0.356873521)	($C_{0}$,	-0.657911379)	($C_{1}$,	-0.942706645)	($C_{2}$,	-1.617170214)	($C_{3}$,	-3.069576045)	($C_{4}$,	-7.13)};

\addplot [color=black, dashed] coordinates {(minus, 0) ($C_{ref}$, 0) ($C_{I}$,	0)	($C_{0}$,	0)	($C_{1}$,	0)	($C_{2}$,	0)	($C_{3}$,	0)	($C_{4}$,	0) (plus, 0)};

\legend{\gls{ilf}-pAware, \gls{ilf}-pUnaware, \gls{ilf}-Adaptive, \gls{pp}-pAware}
%\pgfplotsset{every x tick label/.append style={font=\scriptsize}}
\end{axis}
\end{tikzpicture}
    \caption{Performance evaluation of different \acrshort{ilf} \acrshort{qe} configurations of the \gls{ilf} approach, in terms of \acrshort{bd-br}. The proposed prediction-aware \gls{pp} performance is also presented (black line) in order to magnify the performance drop due to the multiple enhancement effect in the last configurations of the \gls{ilf} approach. All tests are carried out in the \gls{ra} mode (Class C and D, CTC).}
    \label{fig:ilfcurve}
\end{figure}
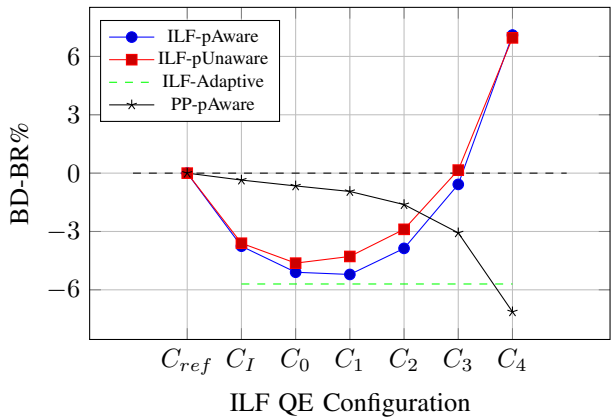

Fig. \ref{fig:ilfcurve} shows the evolution of \gls{ilf} methods in different \acrshort{ilf} \acrshort{qe} configurations. These results are obtained by averaging over class C and D of test sequences. A constant dashed line at \gls{bd-br}=0\% is shown to indicate the border between having compression gain and compression loss.

The first comparison is between the two proposed \gls{ilf} methods presented with red and blue lines, for prediction-unaware and prediction-aware versions, respectively. In these versions the frame-level switch mechanism of the adaptive \acrshort{ilf} method is not used. As can be seen, the proposed prediction-aware \gls{ilf} method outperforms the prediction-unaware method in almost all configurations. 

The effect of the multiple-enhancement can be seen in the shape of both prediction-unaware and prediction-aware \gls{ilf} methods. More precisely, the \gls{bd-br} gain of the \gls{ilf} methods progressively decreases around $C_0$ and $C_1$, and eventually becomes a \gls{bd-br} loss around $C_3$ and $C_4$. 

As the adaptive \gls{ilf} algorithm has the flexibility to apply the \gls{qe} step on any arbitrary frame in the \gls{gop}, the defined \acrshort{ilf} \acrshort{qe} configurations are not applicable to it and its results are presented as a constant green dashed line. The only fair comparison is between the prediction-aware \gls{ilf} (blue line) and the adaptive version. As can be seen, with the adaptive version we can guarantee the highest performance of the non-adaptive version. The reason behind this behaviour is that the adaptive version performs the \gls{mse}-based comparison at each frame ensures that the multiple-enhancement is not going to negatively impact the performance. Therefore, the \gls{qe} task usually stops at the optimum \acrshort{ilf} \acrshort{qe} configuration.

The \gls{pp} algorithm is adapted to the \acrshort{ilf} \acrshort{qe} configurations in order to make a comparison. It is important to note that such comparison with \gls{pp} might not be entirely fair, as the subset of enhanced frames corresponding to given \acrshort{ilf} \acrshort{qe} configurations are not necessarily optimal subsets for \gls{pp}. However, our experiments show that the difference is small enough for drawing a conclusion. The interesting comparison between the \gls{pp} method and the two \gls{ilf} methods is their crossing point. In other words, until around $C_2$, both \gls{ilf} methods are better than the \gls{pp}. However, after this configuration, the \gls{pp} becomes better. The reason is that, in the first part (until $C_2$, the enhancement propagation is causing the \gls{ilf} methods to be better than the \gls{pp}. However, in the second part ($C_3$ and $C_4$), the negative impact of the multiple-enhancement effect entirely compensates the enhancement propagation effect. Therefore, the crossing point of the \gls{pp} method and \gls{ilf} methods shows how the balance between the enhancement propagation and the multiple-enhancement can impact the performance of \gls{ilf} method.

The final observation from Fig. \ref{fig:ilfcurve} is the fact that best performance of the \gls{pp} (at $C_4$) is better than the best performance of the \gls{ilf} (shown with \gls{ilf}-Adaptive line). This is most likely due to the fact that current \gls{ilf} methods are not optimized with end-to-end training to be able to enhance frames in higher temporal layers (\textit{e.g.} at $C_3$ and $C_4$). Improvement of this aspect is left as future work.

%-------------------------------------------------------------------------------------------------
\subsubsection{Performance vs Complexity Trade-off}
In general, by applying the \gls{ilf} approach to the frames in lower temporal layers, the side effect of multiple enhancement can be controlled. For instance, if only first I frame is enhanced and other frames remain intact, a noticeable gain will be obtained compared to the fact that only one frame is enhanced. In other words, by imposing the complexity of enhancing only one frame, we obtain a good portion of performance if we apply \gls{qe} to all frames. To test this effect, we have calculated in table \ref{tab:complexity_ilf} the relative complexity of the configurations introduced in section \ref{sec:ILFgopconfigs}. As can be seen, the encoding and decoding time is increasing as we apply \gls{qe} filter to more temporal levels. 

We can also see from Table \ref{sec:ILFgopconfigs} that, the decoder complexity of \gls{ilf} in the adaptive configuration ($C_{ad}$) varies depending on how many frames have been enhanced at the encoder side. On the other hand, using \gls{pp} does not impose any complexity at the encoder side, however at the decoder side, it dramatically increases the complexity.  

\begin{table*}[h]
    \caption{Relative Complexity of the proposed \acrshort{pp} and \acrshort{ilf}, averaged on \glspl{ctc} \acrshort{qp} range, as in Eq. \eqref{eq:complexity}. Here, $C_i$ refers to the \acrshort{ilf} configurations presented in section \ref{sec:ILFconfigs}. All tests have been carried out with the native \acrshort{ra} coding mode of VTM-10.}
    \centering
    \renewcommand{\arraystretch}{1.15} 
\begin{tabular}{P{0.05\textwidth}|P{0.05\textwidth}|P{0.03\textwidth}P{0.03\textwidth}|P{0.03\textwidth}P{0.03\textwidth}|P{0.03\textwidth}P{0.03\textwidth}|P{0.03\textwidth}P{0.03\textwidth}|P{0.03\textwidth}P{0.03\textwidth}|P{0.03\textwidth}P{0.03\textwidth}|P{0.03\textwidth}P{0.03\textwidth}}
\hline
\hline        

\multirow{2}{*}{Platform}   & \multirow{2}{*}{Class}       & \multicolumn{2}{c|}{$PP$}   & \multicolumn{2}{c|}{$ILF-C_{ad}$} & \multicolumn{2}{c|}{$ILF-{C_I}$} & \multicolumn{2}{c|}{$ILF-{C_0}$} & \multicolumn{2}{c|}{$ILF-{C_1}$} & \multicolumn{2}{c|}{$ILF-{C_2}$} & \multicolumn{2}{c}{$ILF-{C_3}$} \\
\cline{3-16}
        &                               &       ET     &       DT   &        ET     &       DT    &        ET     &       DT    &         ET     &       DT   &         ET     &       DT   &         ET     &       DT   &        ET     &       DT    \\
\hline                   
\multirow{4}{*}{CPU}     & B & - & 96.82   &  1.08  &  37.18  &  1.00  &  3.95    &  1.01  &  8.37  &  1.01  &  14.27  &  1.02  &  26.06  &  1.04  &  48.50 \\
						 & C & - & 329.87  &  1.17  &  78.76  &  1.01  &  12.27  &  1.01  &  26.30  &  1.02  &  46.54  &  1.05  &  87.01  &  1.09  &  167.96 \\ 
						 & D & - & 312.71  &  1.21  &  90.17  &  1.01  &  11.58  &  1.02  &  24.98  &  1.03  &  44.16  &  1.05  &  82.53  &  1.10  &  159.25 \\
\hline
\hline
                          
\multirow{4}{*}{GPU}    & B & - & 25.06  & 1.02 & 4.15  & 1.00 & 1.74  & 1.00 & 2.85  & 1.00 & 4.33  & 1.01 & 7.29  & 1.01  & 12.92 \\
                        & C & - & 14.36  & 1.01 & 2.6  & 1.00 & 1.46  & 1.00 & 2.03  & 1.00 & 2.85  & 1.00 & 4.49  & 1.00  & 7.78  \\
                        & D & - & 13.61  &   1.04  &   3.5  &   1.00  &   1.38  &   1.01  &   1.93  &   1.01  &   2.70  &   1.00  &   4.26  &   1.01  &   7.38   \\

\hline
\hline

\end{tabular}

    \label{tab:complexity_ilf}
\end{table*}

%-------------------------------------------------------------------------------------------------

\subsubsection{Performance comparison with state-of-the-art of ILF} 
Table \ref{tab:stateoftheart_ilf} compares the performance of proposed \gls{qe} methods, when integrated as \gls{ilf}, against some state-of-the-art methods in literature. For this purpose, we chose MFRNet \cite{ma2020mfrnet} and ADCNN \cite{wang2019attention} methods from academic papers in addition to three recent JVET contributions. As the source code of most of these works is not publicly available, the performance metric have been directly extracted from corresponding papers and documents. For representing both prediction-unaware and prediction-aware methods, we used the adaptive \gls{ilf} implementation, described in Section \ref{sec:switchable}, since it provides the highest performance. 

It can be observed that our proposed prediction-aware \gls{ilf} outperforms the proposed prediction-unaware method, by the coding gain of -5.85\% compared to -5.12\%, showing a consistent average \gls{bd-br} gain of about -0.73\%. This result shows once more how the use of the prediction information can further improve the performance of a given \gls{cnn}-based \gls{qe} method.

Furthermore, the proposed prediction-aware method also significantly outperforms the selected papers from the state-of-the-art. It is worth to mention that since the benchmark methods use different network architecture and sometimes different training and test settings, this comparison might not entirely be fair. 
%\hl{How can we mention this?}. 
As future work, more efficient network architectures can be adopted from some of the benchmarks methods and integrated into the proposed prediction-aware framework of this paper. Or inversely, one can implement the prediction-aware aspect of the proposed method on top of the benchmark methods and measure its performance changes.

In another analysis, the \gls{ilf} results of Table \ref{tab:stateoftheart_ilf} can be compared against the \gls{pp} results in Table \ref{tab:stateoftheartpp}. By doing so, it can be observed that the \gls{pp} approach is -1.4\% better than the \gls{ilf} one. This is mainly due to the multiple enhancement effect and the fact the \gls{cnn}-based enhancement in some high temporal layers in the \gls{gop} have been avoided by the adaptive mechanism of the proposed method in order to control the quality degradation.

\begin{table*}[h]
    \caption{ BD-BR comparison of the proposed  \gls{ilf} method against state-of-the-art methods, computed on the \acrshort{ra} mode.}
    \centering
    
\renewcommand{\arraystretch}{1.15} 
\begin{tabular}{P{0.05\textwidth}P{0.1\textwidth}P{0.1\textwidth}P{0.1\textwidth}P{0.08\textwidth}P{0.08\textwidth}P{0.1\textwidth}|P{0.08\textwidth}P{0.08\textwidth}}

\hline
\hline							 

\multirow{3}{*}{Class}	& \multicolumn{6}{c|}{State-of-the-art}					 & \multicolumn{2}{c}{Proposed} \\ 
\cline{2-9} 
						& JVET-O0079 \cite{JVETO0079}       & JVET-T0088 \cite{JVETT0088}& JVET-U0054 \cite{JVETU0054}& MFRNet \cite{ma2020mfrnet}&ADCNN  \cite{wang2019attention}&JVET-T0079 \cite{JVETT0079}         & Pred-unaware          & Pred-aware         \\
						     & VTM-5				              & VTM-9			    & VTM-10				         & VTM-7				   & VTM-4				          & VTM-10				        & VTM-10				    & VTM-10					\\
\hline	                  
B     					&  0.64\%                           & -3.44\%                   & -4.04\%                        & -4.30\%                & -1.53\%                             & -3.25\%                       & -5.12\%               & -5.85\%             \\  
C     					& -1.17\%                           & -3.38\%                   & -4.69\%                        & -3.30\%                & -3.06\%                             & -2.85\%                       & -4.36\%               & -5.13\%             \\  
D     					& -3.13\%                           & -3.48\%                   & -6.20\%                        & -5.50\%                & -3.83\%                             & -3.13\%                       & -5.87\%                & -6.58\%             \\  
\hline                                                                                                                                                       
\textbf{All}            & \textbf{-1.22\%}                  & \textbf{-3.43\%}          & \textbf{-4.98\%}               & \textbf{-4.37\%}       & \textbf{-2.81\%}                    & \textbf{-3.08\%}          & \textbf{-5.12\%}           &\textbf{-5.85\%}          \\ 

\hline                                                                                                                        
\hline
%\makegapedcells
\end{tabular}

    \label{tab:stateoftheart_ilf}
\end{table*}

%-------------------------------------------------------------------------------------------------
%-------------------------------------------------------------------------------------------------
%-------------------------------------------------------------------------------------------------

\section{Conclusion}
\label{sec:conclusion}
In this paper, we have proposed a \gls{cnn}-based \gls{qe} method to address the \acrfull{pp} and \acrfull{ilf} problems in \gls{vvc}. Precisely, a filter which exploits the coding information such as prediction and \gls{qp} is proposed in order to better enhance the quality. These coding information is fed to a proposed \gls{qe} network based on the frame coding type (intra-frame or inter-frame), resulting in several trained models. Depending on the coding type used for a block (e.g inter mode, intra mode or skip mode), a model is selected among three models for the \gls{qe} task. Moreover, in the \gls{ilf} integration, in order to avoid the multiple enhancement issue, we adopt an adaptive framework to skip enhancement of frames posing this problem. Experimental results showed that the proposed \gls{pp}, as well as \gls{ilf} methods, outperform the state-of-art methods in terms of \gls{bd-br}. 

There are rooms for improvement of this work. For instance, the proposed \gls{qe} framework enhances all three components using one single model for a given coding type. However, the content of chroma components has different characteristics which could be taken into account separately, during the training and inference. Moreover, there is plenty of coding information available during the encoding and within the bitstream at the decoder side. This coding information could provide a variety of spatial and temporal features from the video. The promising improvement we achieved from using the prediction signal and \gls{qp} is encouraging to deeper investigate the effect of other coding information. Therefore, as another future work, one can benefit from other coding information for further improving the proposed \gls{qe} framework. Finally, the network can be further simplified by using more efficient architecture in order to decrease the number of parameters, leading to a faster inference. 

\bibliographystyle{ieeetr}
\bibliography{mybib}

\end{document}